\documentclass{aa}  

\usepackage{graphicx}
\usepackage{txfonts}
\usepackage[toc,page]{appendix}
\usepackage{rotating}
\usepackage{geometry}
\usepackage{scalefnt}
\usepackage[colorlinks=true, allcolors=blue]{hyperref}
\begin{document}

   \title{CAPOS: The bulge Cluster APOgee Survey VII. First detailed chemical analysis of NGC 6316}

   \author{Heinz Frelijj\inst{1,2}\thanks{E-mail: h.frelijjrubilar@uandresbello.edu},
          Danilo Gonz\'alez-D\'iaz\inst{2},
          Doug Geisler\inst{3,4}, 
          Sandro Villanova\inst{1},
          Cesar Mu\~{n}oz\inst{3},
          Christian Moni Bidin\inst{2}
          }

   \institute{Universidad Andres Bello, Facultad de Ciencias Exactas, Departamento de F{\'i}sica y Astronom{\'i}a - Instituto de Astrof{\'i}sica, Autopista Concepci\'on-Talcahuano 7100, Talcahuano, Chile\\
   Instituto de Astronomía, Universidad Católica del Norte, Av. Angamos 0610, Antofagasta, Chile.\\
   Departamento de Astronomía, Facultad de Ciencias, Universidad de La Serena. Av. Juan Cisternas 1200, La Serena, Chile.\\
   Universidad de Concepción, Departamento de Astronomía, Casilla 160-C, Concepción, Chile.\\
             }

   \date{Received September 15, 1996; accepted March 16, 1997}

\titlerunning{CAPOS VII: Chemical analysis of NGC 6316}
\authorrunning{Frelijj et al.}
 
  \abstract 
   {The Bulge Globular Cluster NGC 6316 has had few previous chemical studies beyond metallicity. As part of the bulge Cluster APOgee Survey (CAPOS), we can now improve our knowledge of the chemistry and nature of this cluster.
   CAPOS makes use of high-resolution, high signal-to-noise ratio near-infrared spectroscopy, which is capable of penetrating the substantial dust towards the Galactic Bulge - a significant optical obstacle for this cluster due to its high reddening.}
   {We aim to conduct the most robust chemical study to date for NGC 6316 by deriving abundances for a number of elements with a variety of nucleosynthetic origins, most of which have never been studied before in this cluster.}
   {We use the Brussels Automatic Code for Characterizing High accuracy Spectra (BACCHUS) with atmospheric parameters photometrically obtained in order to determine the abundances.}
   {We determined, for the first time, high-resolution spectroscopic abundances for C, N, O, Mg, Al, Si, P, K, Ca, Ti, V, Cr, Mn, Fe, Ni and Ce for this cluster. We obtained a mean $[Fe/H]=-0.87\pm0.02$, finding no indication of an intrinsic metallicity spread. Our metallicity agrees with the most recent values from other studies, revising earlier values that were $\sim$0.5 dex metal-richer. With this new value, this cluster, long believed to be a member of the classical metal-rich group of bulge GCs around -0.5, now falls in the dominant bulge globular cluster peak around $[Fe/H]=-1$. The cluster presents a clear C-N anti-correlation and [$\alpha$/Fe]=0.31$\pm$0.02. Our abundances show similar behaviour to other in-situ globular clusters with comparable metallicity. An isochrone fitting gave us $E(B-V) = 0.71$, a higher value than any other from the literature for this cluster since we also estimated $R_V=2.7$, in good agreement with determinations from other works; $(M-m)_0=15.32\pm0.05$. We derive an overall metallicity $[M/H]=-0.6\pm0.05$, in agreement with our abundance determination.}
    {}

   \maketitle

\section{Introduction}
\label{Introduction}

Globular Clusters (GCs) are among the oldest objects in the universe. This characteristic make GCs very interesting objects to study in the context of galaxy and cluster formation. Because of this, they have been the main target of a vast amount of research, which has yielded unprecedented results in these fields. These include cluster distances, reddening, and ages \citep{Dotter2010,Cohen2012,Vandenberg2013}, luminosity and mass functions \citep{Paust2010}, structural parameters and mass segregation measurements \citep{Goldsbury2013}, mass loss \citep{Caloi2008,Momany2012} and studies in multiple populations \citep[MPs;][]{Carretta2009a,Villanova2013,Piotto2015,Milone2017}, among others. Significant advances have been made in our understanding of GCs and the formation history of the Milky Way. However, a consistent theoretical framework is still lacking, and key questions remain regarding the fundamental nature of GCs and the broader history of our Galaxy.

\begin{table*}[ht]
\centering
\caption{\label{prop} Table with properties of the targets.}
\resizebox{\textwidth}{!}{
 \begin{tabular}{c c c c c c c c c c c c c c} 
 \hline
  APOGEE ID & RA(J2000) & DEC(J2000) & G & ${\rm G_{BP}}$ & ${\rm G_{RP}}$ & J & H & K$_s$ & pmRA & pmDEC & RV & SNR\\
& (deg) & (deg) & & & & & & & (mas$\cdot$yr$^{\text{$-$1}}$) & (mas$\cdot$yr$^{\text{$-$1}}$) & (km$\cdot$s$^{\text{$-$1}}$) & (pixel$^{\text{$-$1}}$)\\
  \hline
2M17163864$-$2809385  & 259.161017 & $-28.160715$ & 15.309 & 16.458 & 14.246 & 12.549 & 11.582 & 11.329 & $-4.90$ & $-4.66$ &  94 &  65\\
2M17165235$-$2809502  & 259.218166 & $-28.163954$ & 15.351 & 16.521 & 14.267 & 12.421 & 11.571 & 11.336 & $-4.81$ & $-4.55$ & 101 &  70\\
2M17163623$-$2808067  & 259.150986 & $-28.135221$ & 14.713 & 15.835 & 13.478 & 11.55  & 10.213 & 10.212 & $-4.94$ & $-4.68$ &  99 &  80\\
2M17163330$-$2808396  & 259.138767 & $-28.144335$ & 14.957 & 16.115 & 13.800 & 11.994 & 10.944 & 10.758 & $-4.80$ & $-4.77$ & 102 &  85\\
2M17164048$-$2808443  & 259.168675 & $-28.145658$ & 15.210 & 16.262 & 14.045 & 12.274 & 11.325 & 11.046 & $-4.99$ & $-4.73$ & 103 &  85\\
2M17164482$-$2808302  & 259.186784 & $-28.141745$ & 15.315 & 16.474 & 14.242 & 12.475 & 11.544 & 11.210 & $-4.96$ & $-4.46$ & 101 &  90\\
2M17163393$-$2811052* & 259.141393 & $-28.184803$ & 14.742 & 15.988 & 13.624 & 11.496 & 10.646 & 10.375 & $-4.91$ & $-4.58$ & 103 & 105\\
2M17163903$-$2807212* & 259.162634 & $-28.122564$ & 14.683 & 16.009 & 13.555 & 11.633 & 10.574 & 10.319 & $-4.73$ & $-4.64$ &  94 & 135\\
2M17163911$-$2804506  & 259.162961 & $-28.080727$ & 14.549 & 15.962 & 13.357 & 11.282 & 10.239 &  9.875 & $-4.83$ & $-4.51$ & 102 & 140\\
2M17163627$-$2807166  & 259.151135 & $-28.121305$ & 14.247 & 15.884 & 13.017 & 10.923 & 9.838  &  9.503 & $-4.98$ & $-4.63$ & 103 & 170\\
\hline
2M17162672$-$2811418  & 259.111349 & $-28.194954$ & 16.483 & 17.576 & 15.454 & 13.796 & 12.909 & 12.763 & $-5.07$ & $-4.68$ &  94 &  30\\
2M17163992$-$2805581  & 259.166366 & $-28.099491$ & 16.636 & 17.611 & 15.478 & 13.834 & 12.943 & 12.749 & $-4.98$ & $-4.67$ &  97 &  30\\
2M17165082$-$2807196  & 259.211757 & $-28.122114$ & 16.280 & 17.245 & 15.278 & 13.660 & 12.855 & 12.573 & $-4.88$ & $-4.42$ &  98 &  35\\
2M17163622$-$2812074  & 259.150947 & $-28.202074$ & 15.717 & 16.789 & 14.691 & 12.976 & 12.165 & 11.894 & $-5.08$ & $-4.54$ & 102 &  40\\
2M17164149$-$2807547  & 259.172901 & $-28.131874$ & 15.539 & 16.584 & 14.478 & 12.839 & 11.959 & 11.648 & $-5.23$ & $-4.51$ &  92 &  45\\
2M17162882$-$2807520  & 259.120092 & $-28.131128$ & 16.138 & 17.257 & 15.104 & 13.393 & 12.534 & 12.278 & $-4.96$ & $-4.76$ &  95 &  50\\
2M17164437$-$2806426  & 259.184902 & $-28.111856$ & 16.061 & 17.137 & 15.050 & 13.406 & 12.567 & 12.324 & $-5.08$ & $-4.63$ & 107 &  55\\
2M17162377$-$2809447  & 259.099078 & $-28.162430$ & 15.327 & 16.483 & 14.254 & 12.540 & 11.602 & 11.339 & $-4.86$ & $-4.68$ & 103 &  55\\
2M17163552$-$2806556  & 259.148019 & $-28.115461$ & 15.473 & 16.573 & 14.422 & 12.753 & 11.818 & 11.609 & $-4.88$ & $-4.84$ & 105 &  55\\
 \hline
\end{tabular}
}
\tablefoot{The top group contains the stars included in the final sample. The bottom group includes stars that are inside the TR of the cluster, match with the PMs, RVs and ASPCAP [Fe/H] of the cluster, and are within the RGB but were ultimately discarded due to their low SNR.\\
The order of the columns is as follows: Star ID, Right Ascension(J2000), Declination(J2000), GAIA DR3 \citep{Gaia2016,Gaia2021} magnitude in G, G$_{\rm BP}$ and G$_{\rm RP}$, 2MASS magnitude in J, H and K$_s$, Absolute Proper Motion in RA and DEC from GAIA DR3, Heliocentric RV from ASPCAP, and SNR. IDs marked with an * indicate spectra not obtained by CAPOS.}
\end{table*} 
 
Given that many questions remain unanswered after studying a large number of primarily halo GCs, a promising direction to focus future studies is to turn to the study of GCs in other Galactic components. The Galaxy possesses no less than three GC systems: the halo, bulge and disk systems \citep[e.g.][]{Massari2019}. Indeed, these latter two contain fully 2/3 of the total number of Galactic GCs \citep{Belokurov2024}. The study of bulge globular clusters (BGCs) in particular opens up the regime of high metallicity to MP studies since they include the highest metallicity of all GCs, and thus help verify metallicity trends so far only hinted at \citep[e.g.][]{Munoz2020}. Nevertheless, most of the Bulge/Disk (B/D) GC systems have not been studied in the detail required to quantify or even explore their MP nature, which demands either deep UV photometry with high spatial resolution, or high resolution high Signal-to-Noise Ratio (SNR) spectroscopy of large samples. 
Unfortunately, due to the vast amount of extinction in the plane and towards the centre of the Galaxy \citep[e.g.][]{Gonzalez2012}, UV observations of the B/D GC systems are severely limited, so that the HST UV Legacy Survey \citep{Piotto2015} includes only a very small number of these objects. Optical spectroscopy is similarly limited. However, the high reddening problem that previously prevented us from studying the B/D can be quite successfully addressed with the help of near-infrared (NIR) instruments. Surveys like the VISTA Variables in the Via Lactea (VVV) \citep{Minniti2010} with its eXtension (VVVX)\citep{Saito2024}, and the Apache Point Observatory Galactic Evolution Experiment of the Sloan Digital Sky Survey III (SDSS-III APOGEE) \citep{Majewski2017} with its successor (SDSS-IV APOGEE-2) capitalize on this capability to great success. However, despite these large surveys, the VVV survey lacks UV observations while the SDSS-IV survey did not prioritize B/D GCs. \citet{Meszaros2020} present the SDSS-IV sample of 44 GCs, of which only 2 are bona fide BGCs according to \citet{Massari2019} and have a large enough sample of well-observed members and low enough reddening to qualify for their study. This is <4\% of the total number of BGCs known, a disconcertingly low value. 

Despite the existence of multiple individual studies of BGCs \citep[e.g.][among others]{Origlia2005,Valenti2011,Villanova2017,Munoz2017}, a homogeneous study, involving a large number of BGCs is still lacking. In order to take full advantage of the wealth of astrophysical detail these key objects can provide, as complete a sample as possible is essential. To this end, "the bulge Cluster APOgee Survey" (CAPOS) \citep{Geisler2021} was created. This survey was designed to obtain detailed abundances and kinematics for as complete a sample of bona fide BGCs as possible using the unique advantages of APOGEE in order to exploit their extraordinary Galactic archaeology attributes. In addition to providing reliable multi-element abundances for a number of B/D GCs (many for the first time), this ongoing project has published a series of important results: e.g. confirmation of the scenario that Main Bulge and Main Disk GCs, formed in situ, have [Si/Fe] abundances slightly higher than their accreted counterparts at the same metallicity \citep[CAPOS I Paper]{Geisler2021}; the existence of the characteristic N-C anti-correlation, and Al-N correlation for the first time in FSR 1758 \citep[CAPOS II Paper]{Romero-Colmenares2021}; a Ce-N correlation in Tonantzintla 2 \citep[CAPOS III Paper]{Fernandez2022}; a N-C anti-correlation for the first time in NGC 6558 \citep[CAPOS IV Paper]{GonzalezDiaz2023}, and detailed studies of HP 1 \citep[CAPOS V Paper]{Henao2025} and NGC 6569 \citep[CAPOS VI Paper]{Barrera2025}. These studies, which are providing valuable information about B/D GCs, follow a coordinated methodology that provides homogeneous studies, allowing a uniform analysis of the data and results.

 Although there is already an APOGEE pipeline that derives chemical abundances and other parameters like radial velocity for BGCs, namely the APOGEE Stellar Parameters and Chemical Abundance Pipeline (ASPCAP, \citet{Garcia2016}), it is known that the pipeline does not perform optimally for stars with extreme types of “non-standard” elemental abundance patterns, namely so-called Second Population (2P) GC stars - those that have been enriched by some as yet-to-be-determined process \citep{Holtzman2015,Jonsson2018}. \citet{Fernandez2020} found significant differences in certain abundance ratios compared to ASPCAP, in the range of $\sim 0.1-0.75$ dex higher than ASPCAP values for [N, O, Al, Si, Ce/Fe] to $\sim 0.05-0.4$ dex lower for [O, Mg/Fe] in 2P stars. CAPOS I showed that even the ASPCAP Fe abundances of 2P stars deviates from that of their first population (1P) cousins, with the former offset to somewhat higher metallicity values within a given GC. This was strongly corroborated by Geisler et al. (in prep) recent analysis of APOGEE results for all CAPOS GCs. Hence, different abundance determination techniques are necessary for 2P stars at least, and preferably one should also perform the same analysis on 1P stars in order to ensure homogeneity.
 
 In this paper we focus on a relatively poorly studied BGC, NGC 6316. This cluster was observed as part of CAPOS along with another BGC, NGC 6304, in the same APOGEE field, in order to maximize the number of CAPOS clusters. Another SDSS-IV program also observed stars in this same general area, and some of those stars also ended up being NGC 6316 members. \citet{Meszaros2020} notes the existence of only a single star of NGC 6316 in their large-scale study of GCs with APOGEE, but did not publish any abundances. The total number of APOGEE members studied here, 10, provides a good sample for not only deriving mean abundances but also investigating MPs. 
 
 NGC 6316 is regarded as one of the classic metal-rich BGCs, with a metallicity of $-$0.45 listed by \citet[2010 update, hereafter H10]{Harris1996}. However, due to the extinction problem described above, no previous high resolution spectroscopic study has been performed on it, and it is not part of the HST UV Legacy survey, so nothing is known of its MP nature. Its metallicity has only been estimated via relatively uncertain photometric or low resolution spectroscopic means, e.g. \citet[hereafter ZW84]{Zinn1984}(Pseudo-equivalent width transformed to $Q_{39}$: $-0.47\pm0.15$), \citet[hereafter AZ88]{Armandroff1988}(Calcium triplet: $-0.62$; averaged with ZW84: $-0.55$), \citet[hereafter VFO07]{Valenti2007}(mid-resolution photometry: $-0.58$), \citet[hereafter C09]{Carretta2009c}(UVES recalibration of the 1984 $Q_{39}$: $-0.36\pm0.14$), \citet{Conroy2018}(stellar population models: $-0.87\pm0.03$). The best spectroscopic study, until now, is the Calcium triplet study of \citet{Vasquez2018}, who found a mean metallicity of $-0.35\pm0.14$, $-0.37\pm0.17$ and $-0.63\pm0.21$ using [Fe/H]$-\langle W'\rangle$ calibrations from \citet{Dias2016}, \citet{Saviane2012} and \citet{Vasquez2015}, respectively, from only 3 stars, in good agreement with the H10 catalogue. Most recently, \citet[hereafter D23]{Deras2023} used deep HST photometry to estimate the age and metallicity, deriving [Fe/H] $\sim$ $-$0.9 and 13.1 $\pm$ 0.5 Gyr, making it "another extremely old relic of the assembly history of the Galaxy" according to them, but they lament the lack of a high-quality metallicity determination for this GC, which of course is critical for deriving an accurate age. Note that their value is about 0.5 dex lower than previous estimates, further demanding a more robust metallicity derivation for this cluster. 

According to the Baumgardt database\footnote{https://people.smp.uq.edu.au/HolgerBaumgardt/globular/} (and references therein), NGC 6316 lies at J2000 coordinates RA: 259.15542 deg and DEC: $-$28.14011 deg, it is located at a heliocentric distance of $R_{\odot} = 11.15 \pm 0.39$ kpc and Galactocentric distance of 3.16 $\pm$ 0.36 kpc, has mean absolute proper motions (PMs) of pmRA = $-$4.974 $\pm$ 0.028 mas$\cdot$yr$^{-1}$ and pmDEC = $-$4.605 $\pm$ 0.027 mas$\cdot$yr$^{-1}$, a mean heliocentric radial velocity (RV) of $R_v = 99.65 \pm 0.84$km$\cdot$sec$^{-1}$, a mass of $3.5\cdot10^5$M$_\odot$, and a tidal radius (TR) $r_t = 47.48$ pc. D23 provides a foreground reddening of E(B-V) $= 0.64 \pm 0.01$.

\citet{Massari2019} classify it as a Main Bulge GC, \citet{Perez2020} as Thick Disk, \citet{Callingham2022} as most likely originating from Kraken, and both \citet{Belokurov2024} and \citet{Chen2024} as in situ. Geisler et al. (in prep) give NGC 6316 an 81\% probability of being an in situ GC and a 51\% probability of being a BGC as opposed to Disk GC, and finally classify it as a bona fide BGC.

 Here we present the data used, how it was obtained, reduced, and members selected (Section \ref{Data}). The derivation of atmospheric parameters and the methodology used to obtain the abundances appear in section \ref{Metodology}. The description of the abundances, analysis, comparison with the literature and analysis of its in-situ nature is given in section \ref{Results}, while Section \ref{Conclusions} contains the summary of the paper.\\

 \begin{figure*}
\centering
  	\includegraphics[width=0.8\columnwidth]{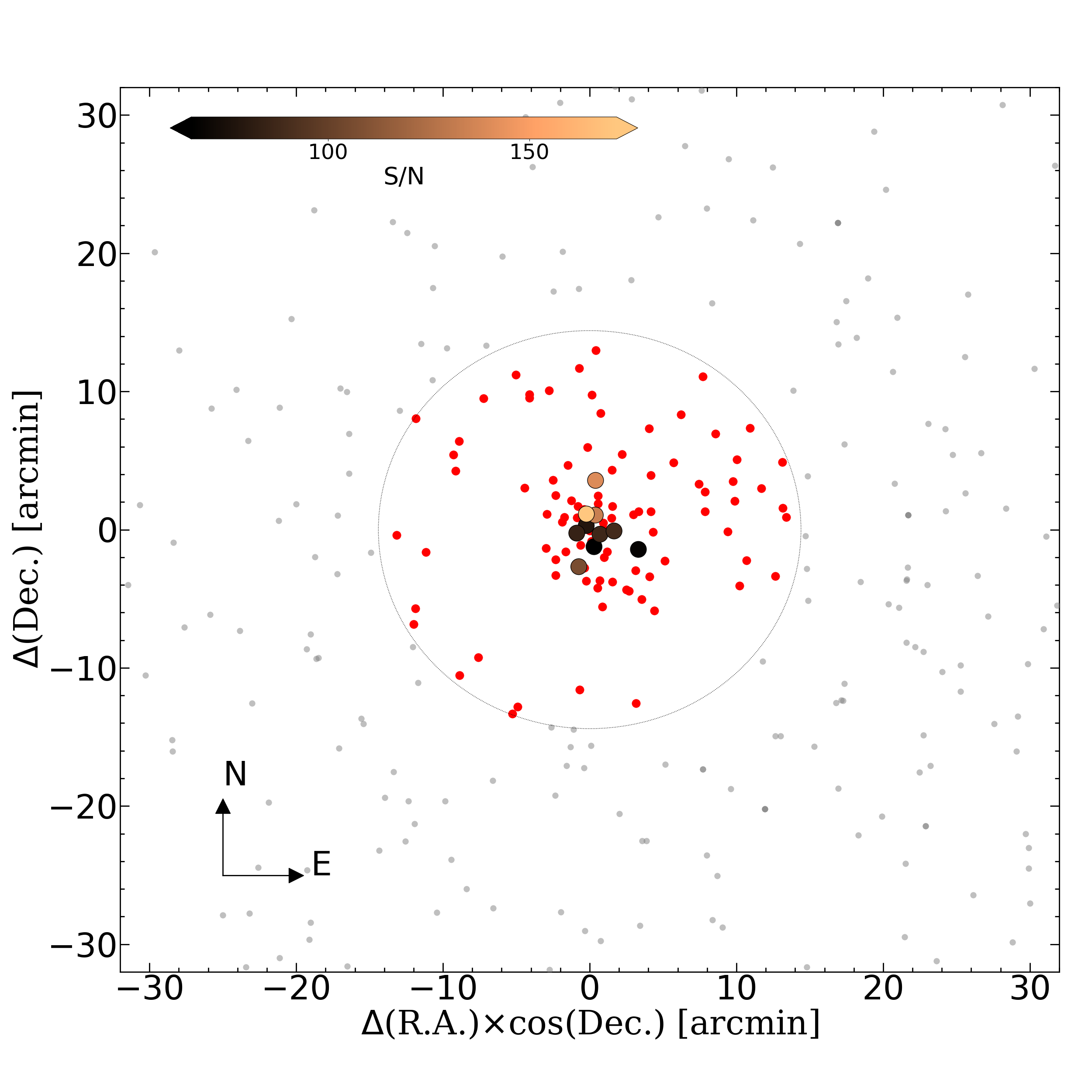}
  	\includegraphics[width=0.8\columnwidth]{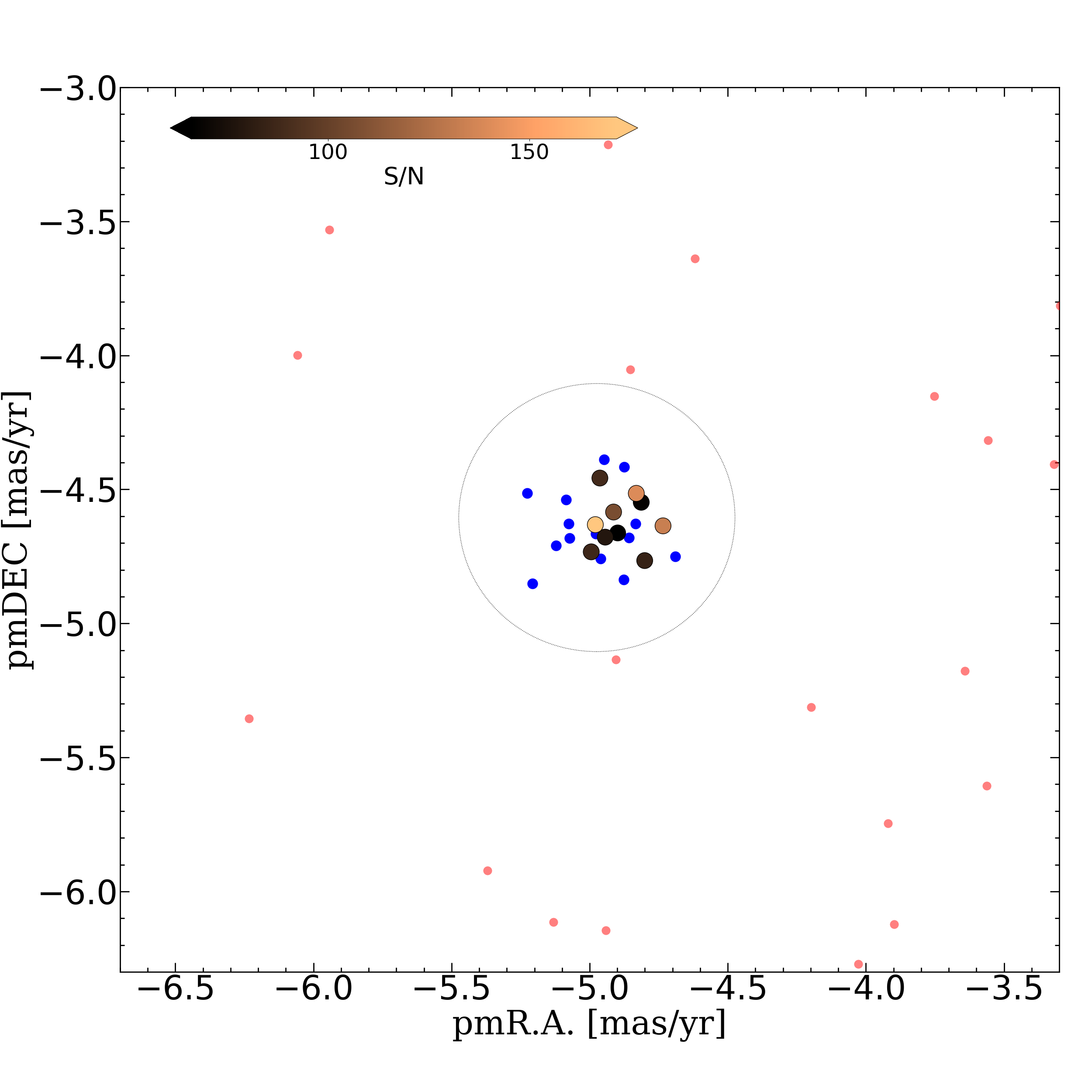}
 	\includegraphics[width=0.8\columnwidth]{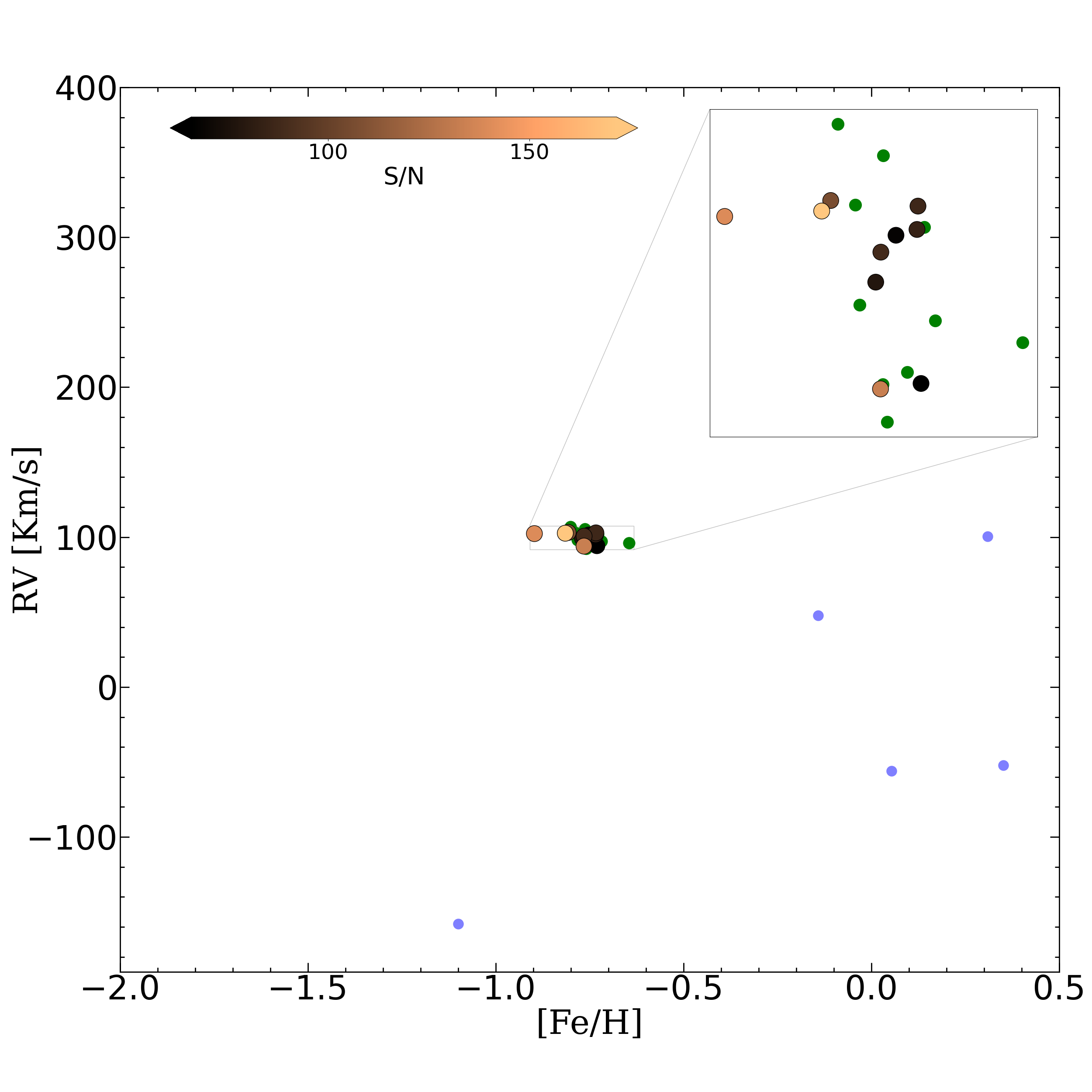}
	\includegraphics[width=0.8\columnwidth]{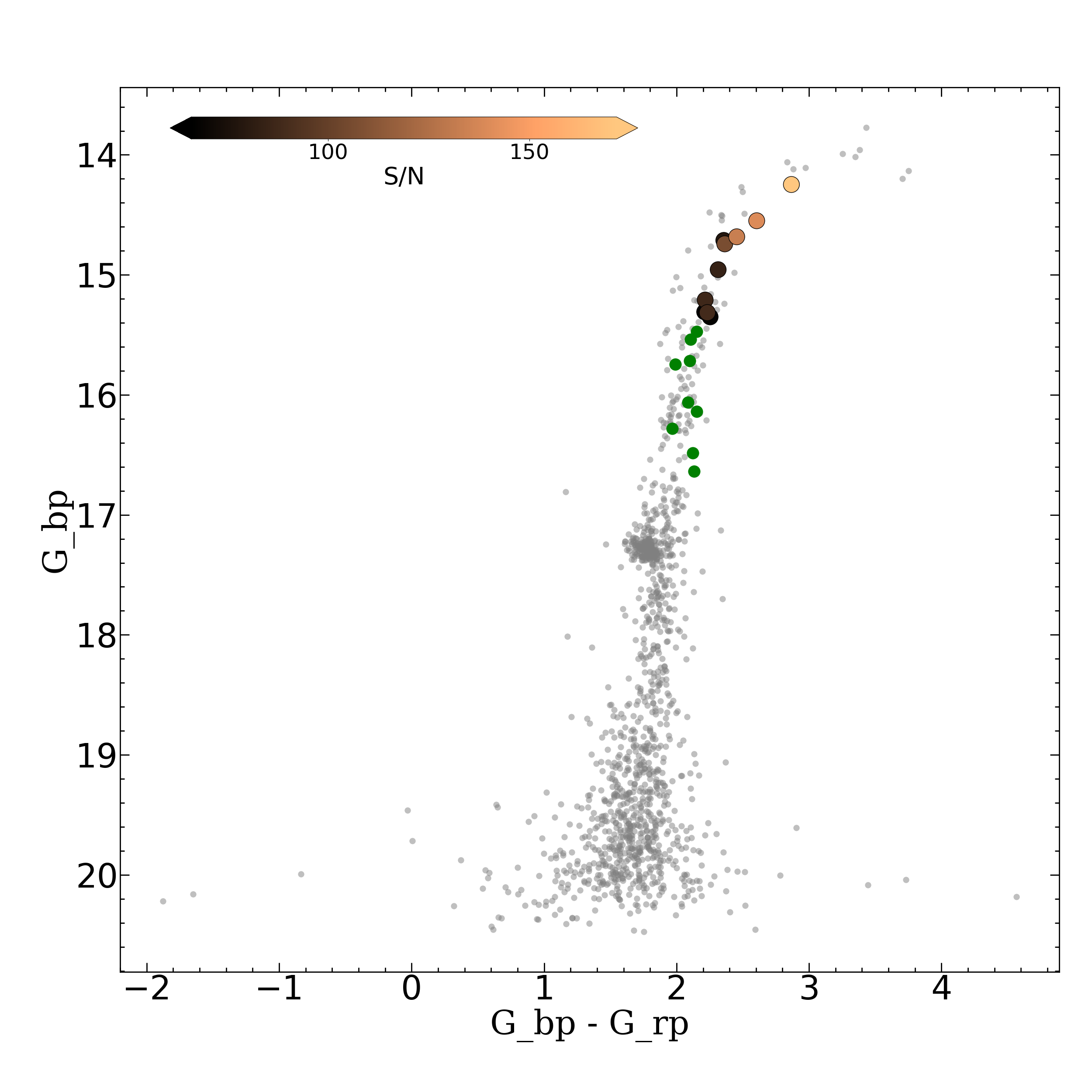}
    \caption{Selection of targets. \textit{Top left:} Selection by position within the cluster tidal radius (TR). Gray dots represent field stars, while red dots represent all stars inside the TR. \textit{Top Right:} Selection by proper motions. Blue stars represent the stars within the fixed tolerance of PMs, while light red dots represent the stars selected in the top left plot by TR but do not fall within the PM limits. \textit{Bottom Left:} Selection by radial velocity and metallicity. Light blue dots represent stars selected in the top right panel by PMs, while green and other colours represent stars selected by their RVs and ASPCAP metallicity. \textit{Bottom right:} Selection by position in the RGB and SNR: Green points represent stars that passed all the above criteria but lack the minimal selected SNR, while gray points represent other cluster stars.\\
        The final sample is depicted in all plots as bigger circles with a colour code depending on their SNR.}
 	\label{selec}

\end{figure*}

\section{Data and target selection}
\label{Data}

 The data consists of high-resolution spectra (R $\sim$ 22500), centred in the near-infrared (H-Band), ranging from 1.51 $\mu$m to 1.70 $\mu$m observed through the APOGEE spectrograph \citep{Wilson2019} on the 2.5m Irénée du Pont telescope at Las Campanas Observatory \citep{Bowen1973}, used by the APOGEE survey to observe the Southern Hemisphere \citep[APOGEE-2S,][]{Santana2021}. 

 NGC 6316 was observed as part of the CAPOS survey \citep{Geisler2021} through the Contributed APOGEE-2S CNTAC program CN2019B. Since the observations were obtained prior to the release of Gaia DR2 data, targets were based on existing position, colour-magnitude and radial velocity information.
 
The final cluster members were selected from the final version of the APOGEE-2 catalogue published in December 2021 as part of the Data Release 17 of the Sloan Digital Sky Survey (SDSS) \citep{Abdurro'uf2022}, which is available in the SDSS IV Science Archive Server\footnote{https://www.sdss4.org/dr17/irspec/spectro\_data/}. 

Based on the information given in this catalogue, we selected observed targets as potential members of NGC 6316 through 5 criteria:\\
\newline
 - Using the central RA/DEC coordinates, distance and physical tidal radius TR of the cluster from Baumgardt database, we determined the angular size of the TR ($\sim$0.24 deg) to select all the observed APOGEE stars inside it (Red dots in the top left panel of Figure \ref{selec})\\
 - We took the mean PMs values from GAIA DR3, attached to the APOGEE catalogue, and selected all the stars inside a radius of 0.5 mas$\cdot$yr$^{-1}$ from the mean value given by Baumgardt database. (Blue dots in the top right panel of Figure \ref{selec})\\
 - Taking into account the mean RV and [Fe/H] of the cluster given by Baumgardt database and D23, we selected a range of stars whose RVs, determined by the APOGEE survey, were around this value, ranging from $92-106$ km$\cdot$sec$^{-1}$, and whose metallicities [Fe/H] from ASPCAP were close to the value given by D23 for the cluster. (Green dots and other non blue circles in the bottom left panel of Figure \ref{selec})\\
 - A colour-magnitude diagram with GAIA DR3 photometry in $G_{bp} - G_{rp}$ vs $G_{bp}$ was made to ensure that all the potential targets lay along the main Red Giant Branch (RGB). (coloured points in bottom right panel of Figure \ref{selec})\\
 - Finally, based on the minimum SNR suggested by APOGEE (SNR $\sim$ 70)\footnote{https://www.sdss4.org/dr17/irspec/abundances/}, 10 out of the 19 potential targets were included in the final membership selection. 
 
A cross-match of our 10 targets with the \citet{Vasiliev2021} catalogue of NGC 6316 determined that, according to the criteria used in their study, each star has a probability of over 80\% of being part of the cluster, adding further reliability that our membership selection is very robust.

 Table \ref{prop} lists the member parameters. These selections led to the previously mentioned total of 10 targets, of which 8 were observed as part of the CAPOS survey \citep{Geisler2021}, and the remaining 2 were observed under the main APOGEE survey \citep{Majewski2017}. The spectra of these 10 targets were downloaded from the SDSS IV Science Archive Server\footnote{\scalefont{0.945}{https://data.sdss.org/sas/dr17/apogee/spectro/aspcap/dr17/synspec\textunderscore rev1/}}. All the obtained spectra had already undergone reduction using the APOGEE-2 data reduction pipeline as described in \citet{Nidever2015}.

\begin{table*}
\caption{\label{abus} Table with chemical abundances and their respective weighted error.}
\centering
\resizebox{0.7\textwidth}{!}{
\hspace*{-5cm}
\begin{tabular}{c c c c c c c c c c c} 
\hline
ID                   &  ${\rm T_{eff}}$ &       log g      &    ${\rm v_t}$   &       [Fe/H]     &      [C/Fe]      &       [N/Fe]     &       [O/Fe]     &      [Mg/Fe]     &      [Al/Fe]     &      [Si/Fe]     \\
2M17163864$-$2809385 &       4119       &       1.19       &       2.49      & $-0.88 \pm 0.05$ & $+0.09 \pm 0.13$ & $+0.73 \pm 0.23$ & $+0.48 \pm 0.23$ & $+0.27 \pm 0.07$ & $+0.29 \pm 0.08$ & $+0.34 \pm 0.05$ \\
2M17165235$-$2809502 &       4118       &       1.19       &       1.67      & $-0.74 \pm 0.06$ & $-0.03 \pm 0.15$ & $+0.10 \pm 0.19$ & $+0.44 \pm 0.20$ & $+0.39 \pm 0.07$ & $+0.39 \pm 0.09$ & $+0.34 \pm 0.05$ \\
2M17163623$-$2808067 &       3842       &       0.68       &       2.95      & $-0.86 \pm 0.10$ & $-0.65 \pm 0.29$ & $+0.93 \pm 0.60$ & $-0.13 \pm 0.42$ & $+0.03 \pm 0.12$ & $+0.27 \pm 0.12$ & $+0.36 \pm 0.09$ \\
2M17163330$-$2808396 &       3979       &       0.93       &       2.24      & $-0.84 \pm 0.09$ & $-0.71 \pm 0.20$ & $+1.40 \pm 0.73$ & $+0.01 \pm 0.28$ & $+0.34 \pm 0.08$ & $+0.16 \pm 0.16$ & $+0.28 \pm 0.09$ \\
2M17164048$-$2808443 &       4051       &       1.06       &       2.17      & $-0.85 \pm 0.07$ & $-0.29 \pm 0.13$ & $+0.93 \pm 0.24$ & $+0.24 \pm 0.24$ & $+0.32 \pm 0.05$ & $+0.35 \pm 0.16$ & $+0.30 \pm 0.04$ \\
2M17164482$-$2808302 &       4083       &       1.12       &       1.69      & $-0.76 \pm 0.08$ & $-0.05 \pm 0.12$ & $+0.11 \pm 0.32$ & $+0.38 \pm 0.21$ & $+0.39 \pm 0.06$ & $+0.37 \pm 0.09$ & $+0.37 \pm 0.04$ \\
2M17163393$-$2811052 &       3884       &       0.76       &       2.70      & $-1.00 \pm 0.05$ & $-0.20 \pm 0.12$ & $+1.11 \pm 0.34$ & $+0.27 \pm 0.20$ & $+0.35 \pm 0.05$ & $+0.09 \pm 0.10$ & $+0.34 \pm 0.07$ \\
2M17163903$-$2807212 &       3871       &       0.73       &       2.80      & $-0.97 \pm 0.08$ & $-0.09 \pm 0.15$ & $+0.91 \pm 0.33$ & $+0.33 \pm 0.20$ & $+0.16 \pm 0.03$ & $+0.29 \pm 0.13$ & $+0.36 \pm 0.06$ \\
2M17163911$-$2804506 &       3754       &       0.52       &       2.23      & $-0.85 \pm 0.07$ & $-0.10 \pm 0.16$ & $+0.42 \pm 0.33$ & $+0.29 \pm 0.25$ & $+0.30 \pm 0.06$ & $-0.09 \pm 0.19$ & $+0.46 \pm 0.07$ \\
2M17163627$-$2807166 &       3668       &       0.36       &       2.72      & $-0.93 \pm 0.10$ & $-0.09 \pm 0.10$ & $+0.67 \pm 0.24$ & $+0.33 \pm 0.19$ & $+0.32 \pm 0.03$ & $+0.14 \pm 0.08$ & $+0.41 \pm 0.07$ \\
\hline
Mean                 &                  &                  &                  & $-0.87 \pm 0.02$ & $-0.21 \pm 0.05$ & $+0.73 \pm 0.12$ & $+0.26 \pm 0.08$ & $+0.29 \pm 0.02$ & $+0.23 \pm 0.04$ & $+0.36 \pm 0.02$ \\
$\sigma$             &                  &                  &                  &       0.08       &       0.25       &       0.40       &       0.18       &       0.11       &       0.14       &       0.05       \\
\hline
\hline
ID                   &   [$\alpha$/Fe]    &      [P/Fe]      &      [K/Fe]      &     [Ca/Fe]      &      [Ti/Fe]     &      [V/Fe]      &      [Cr/Fe]     &      [Mn/Fe]     &      [Ni/Fe]     &      [Ce/Fe]     \\
2M17163864$-$2809385 & $  +0.35 \pm 0.05 $ &        ···       &$ +0.29 \pm 0.06$ & $+0.36 \pm 0.08$ & $+0.43 \pm 0.11$ & $+0.23 \pm 0.16$ & $+0.03 \pm 0.06$ & $+0.01 \pm 0.02$ & $+0.15 \pm 0.06$ & $+0.47 \pm 0.18$ \\
2M17165235$-$2809502 & $  +0.26 \pm 0.06 $ &        ···       &$ +0.21 \pm 0.07$ & $+0.18 \pm 0.11$ & $+0.25 \pm 0.39$ & $+0.16 \pm 0.12$ & $-0.16 \pm 0.15$ &        ···       & $+0.11 \pm 0.07$ &        ···       \\
2M17163623$-$2808067 & $  +0.27 \pm 0.08 $ &        ···       &$ +0.14 \pm 0.07$ & $+0.17 \pm 0.12$ & $+0.01 \pm 0.14$ & $-0.04 \pm 0.13$ & $+0.03 \pm 0.11$ & $+0.08 \pm 0.06$ & $+0.15 \pm 0.12$ &        ···       \\
2M17163330$-$2808396 & $  +0.30 \pm 0.07 $ &        ···       &$ +0.25 \pm 0.13$ & $+0.32 \pm 0.11$ & $+0.18 \pm 0.14$ & $+0.27 \pm 0.15$ & $+0.15 \pm 0.22$ & $+0.10 \pm 0.05$ & $-0.01 \pm 0.07$ & $+0.20 \pm 0.14$ \\
2M17164048$-$2808443 & $  +0.29 \pm 0.04 $ &        ···       &$ +0.23 \pm 0.11$ & $+0.28 \pm 0.08$ & $+0.17 \pm 0.12$ & $+0.32 \pm 0.15$ & $-0.04 \pm 0.09$ & $-0.02 \pm 0.07$ & $+0.11 \pm 0.08$ & $+0.04 \pm 0.14$ \\
2M17164482$-$2808302 & $  +0.34 \pm 0.05 $ & $+0.50 \pm 0.16$ &$ +0.20 \pm 0.07$ & $+0.30 \pm 0.09$ & $+0.26 \pm 0.11$ & $+0.24 \pm 0.11$ & $+0.08 \pm 0.13$ & $-0.19 \pm 0.06$ & $+0.02 \pm 0.04$ & $+0.14 \pm 0.14$ \\
2M17163393$-$2811052 & $  +0.35 \pm 0.05 $ &        ···       &$ +0.42 \pm 0.05$ & $+0.36 \pm 0.07$ & $+0.22 \pm 0.12$ & $+0.26 \pm 0.07$ & $+0.19 \pm 0.06$ & $+0.07 \pm 0.02$ & $+0.14 \pm 0.05$ & $+0.17 \pm 0.16$ \\
2M17163903$-$2807212 & $  +0.35 \pm 0.05 $ & $+0.62 \pm 0.09$ &$ +0.16 \pm 0.09$ & $+0.33 \pm 0.08$ & $+0.24 \pm 0.11$ & $+0.31 \pm 0.12$ & $+0.14 \pm 0.05$ & $+0.06 \pm 0.22$ & $+0.14 \pm 0.05$ & $+0.23 \pm 0.15$ \\
2M17163911$-$2804506 & $  +0.24 \pm 0.06 $ & $+0.46 \pm 0.13$ &$ +0.05 \pm 0.06$ & $+0.02 \pm 0.09$ & $+0.02 \pm 0.11$ & $+0.09 \pm 0.15$ & $-0.15 \pm 0.05$ & $-0.14 \pm 0.04$ & $+0.13 \pm 0.11$ & $-0.05 \pm 0.15$ \\
2M17163627$-$2807166 & $  +0.35 \pm 0.06 $ & $+0.37 \pm 0.13$ &$ +0.24 \pm 0.09$ & $+0.29 \pm 0.09$ & $+0.19 \pm 0.08$ & $+0.32 \pm 0.10$ & $+0.06 \pm 0.10$ & $+0.07 \pm 0.12$ & $+0.09 \pm 0.07$ & $-0.02 \pm 0.10$ \\
\hline
Mean                 & $  +0.31 \pm 0.02 $ & $+0.49 \pm 0.06$ & $+0.22 \pm 0.03$ & $+0.26 \pm 0.03$ & $+0.20 \pm 0.05$ & $+0.22 \pm 0.04$ & $+0.03 \pm 0.04$ & $+0.004\pm 0.03$ & $+0.10 \pm 0.02$ & $+0.15 \pm 0.05$ \\
$\sigma$             &        0.08       &       0.09       &       0.09       &       0.10       &       0.11       &       0.11       &       0.11       &       0.10       &       0.05       &       0.15       \\
\hline
\hline
\end{tabular}
\hspace*{-5cm}
}
\tablefoot{The mean abundance with its associated error, over its associated standard deviation are shown at the bottom of each abundance. The Solar abundances used are from \citet{Asplund2005}.}
\end{table*}

\section{Methodology}
\label{Metodology}

\subsection{Atmospheric parameters}

The atmospheric parameters of our targets were obtained through an interactive procedure based on the GAIA DR3 G, G$_{bp}$, G$_{rp}$ and 2MASS J, H, K$_s$ photometry of the cluster. During this procedure, the effective temperature (T$_{\text{eff}}$) and gravity (log g) of each RGB star (including our targets) were obtained, and at the same time, CMDs were corrected for differential reddening. First, we fitted a PARSEC isochrone \citep{Bressan2012} to the RGB and Red Clump (RC), assuming an age of 13.0 Gyrs. We accounted for reddening by applying the Cardelli et al. (1989) reddening law to the isochrone. The visual absorption A$_V$, the R$_V$ parameter, the intrinsic distance modulus (m-M)$_0$ and the global metallicity [M/H] were determined by simultaneously fitting the RGB, RC and RGB-tip in the K$_s$ vs. G$_{bp}$ - K$_s$, G vs. G$_{bp}$ - G$_{rp}$, G$_{bp}$ vs. G$_{bp}$ - G$_{rp}$ and K$_s$ vs. J - K$_s$ CMDs. The T$_{\text{eff}}$ and log g of each RGB star were then determined as those corresponding to the point on the isochrone where the $K_s$ magnitude matches that of the star (we avoided using RC stars for this step). We used the K$_s$ magnitude because it is the least affected by reddening and, consequently, by differential reddening. Having the temperature we obtained the intrinsic G$_{bp}$ - K$_s$ colour of each star from the colour-temperature relation of the RGB part of the isochrone and, by subtracting the mean reddening obtained from the isochrone fitting, also the differential reddening at the position of each star. Finally, for each star we selected the 4 closest neighbours (5 stars in total) and corrected its G, G$_{bp}$, G$_{rp}$, J, H, K$_s$ magnitudes using the mean differential reddening of the 5 stars. We used the G$_{bp}$ - K$_s$ colour because it is the most sensitive to any reddening variation. This procedure was iterated until any improvement of the CMDs was negligible.

We obtained (m-M)$_0$ = 15.32 $\pm$ 0.05, A$_V$ = 1.91 $\pm$ 0.05 and R$_V$ = 2.7 $\pm$ 0.1, resulting in E(B-V) = 0.71, close to the D23 value of 0.64 but substantially higher than the 0.56 value obtained by VFO07, and the H10 value of 0.54. Indeed, the value is an average between \citet{Zinn1985} and \citet{Reed1988}, who obtained E(B-V) values of 0.47 and 0.61, respectively, with the latter supporting a higher reddening. Our reddening is higher than any other from the literature for this cluster since we estimated R$_V$ = 2.7, lower than the value usually assumed (R$_V$ = 3.1) but in excellent agreement with \citet{Nataf2013}, who determined R$_V$ $\sim$ 2.5 with a dispersion $\sigma_{RV} \sim$ 0.2 toward the inner Galaxy. We also obtained [M/H] = $-$0.6 $\pm$ 0.05, a value higher than the iron content of the cluster ([Fe/H] = $-$0.9) given by D23 (and corroborated by us). This is not surprising since globular clusters are usually $\alpha$-enhanced. Also, the isochrone fitting gave an age estimation of 12 Gyrs, however, we cannot strongly rely on this estimate since no turn off point (TO) is visible in the Gaia CMDs. However, the deep HST data of D23 do indeed reveal the TO and an age of 13.1 $\pm$ 0.5 Gyr.

\subsection{Chemical abundances}
The method used to derive the elemental abundances was a local thermodynamics equilibrium (LTE) analysis using all the 10 targets with the Brussels Automatic Code for Characterizing High accuracy Spectra: BACCHUS \citep{Masseron2016}), which relies on the radiative code Turbospectrum \citep{Alvarez1998,Plez2012} and the MARCS model atmosphere grid \citet{Gustafsson2008}.
With the effective temperature and surface gravity photometrically derived and an initial metallicity estimate from D23 we allowed BACCHUS to determine $v_t$ and the convolution. The $v_t$ is obtained by minimizing the trend of Fe abundances against their reduced equivalent width, while the convolution parameter accounts for the total effect of instrument resolution, macro-turbulence, and $v_t sin(i)$ on the line broadening \citep{Hawkins2016}.

\begin{figure}
\centering
  	\includegraphics[width=0.45\columnwidth]{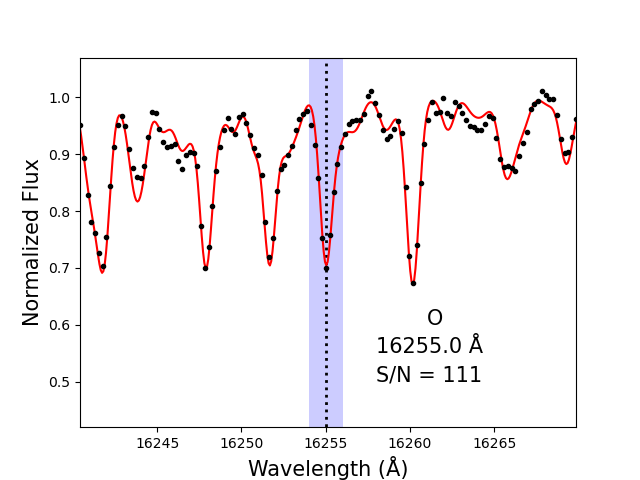}
  	\includegraphics[width=0.45\columnwidth]{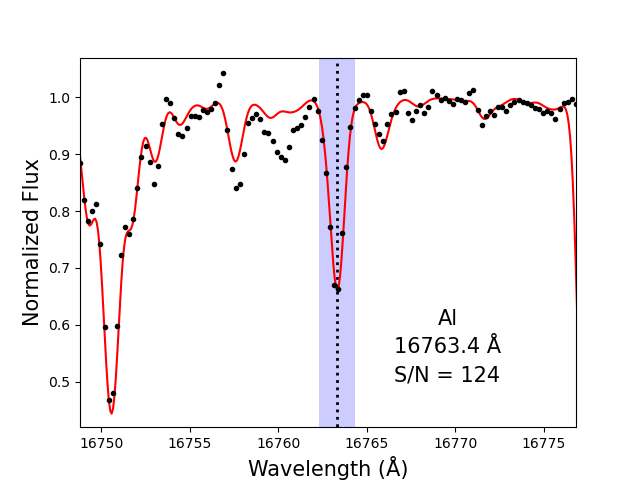}
 	\includegraphics[width=0.45\columnwidth]{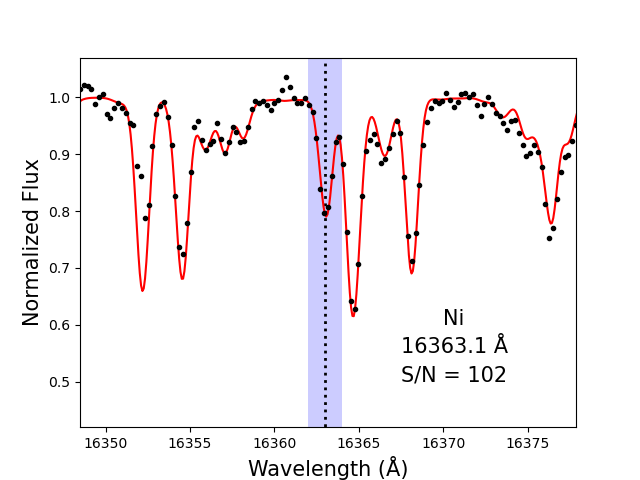}
    \caption{Line fitting showing the quality of the fit for Oxygen, Aluminium and Nickel belonging to the star 2M17163903-2807212}
 	\label{lines}
\end{figure}

We then computed the elemental abundances for C, N, O, Mg, Al, Si, P, K, Ca, Ti, V, Cr, Mn, Fe, Ni and Ce following the same methodology as described in \citep{Hawkins2016}, briefly outlined here: First, to fit the local continuum, we performed a synthesis using the SDSS-III APOGEE spectral line list for H-band spectroscopy \citep{Shetrone2015} updated with the Nd II and Ce II lines from \citet{Hasselquist2016} and \citet{Cunha2017}, respectively. After that, we removed cosmic rays and telluric lines, and estimated the local SNR. Finally, before determining the abundances, a series of flux points contributing to a given absorption line was automatically selected.

\begin{figure*}
\centering
  	\includegraphics[width=1\linewidth]{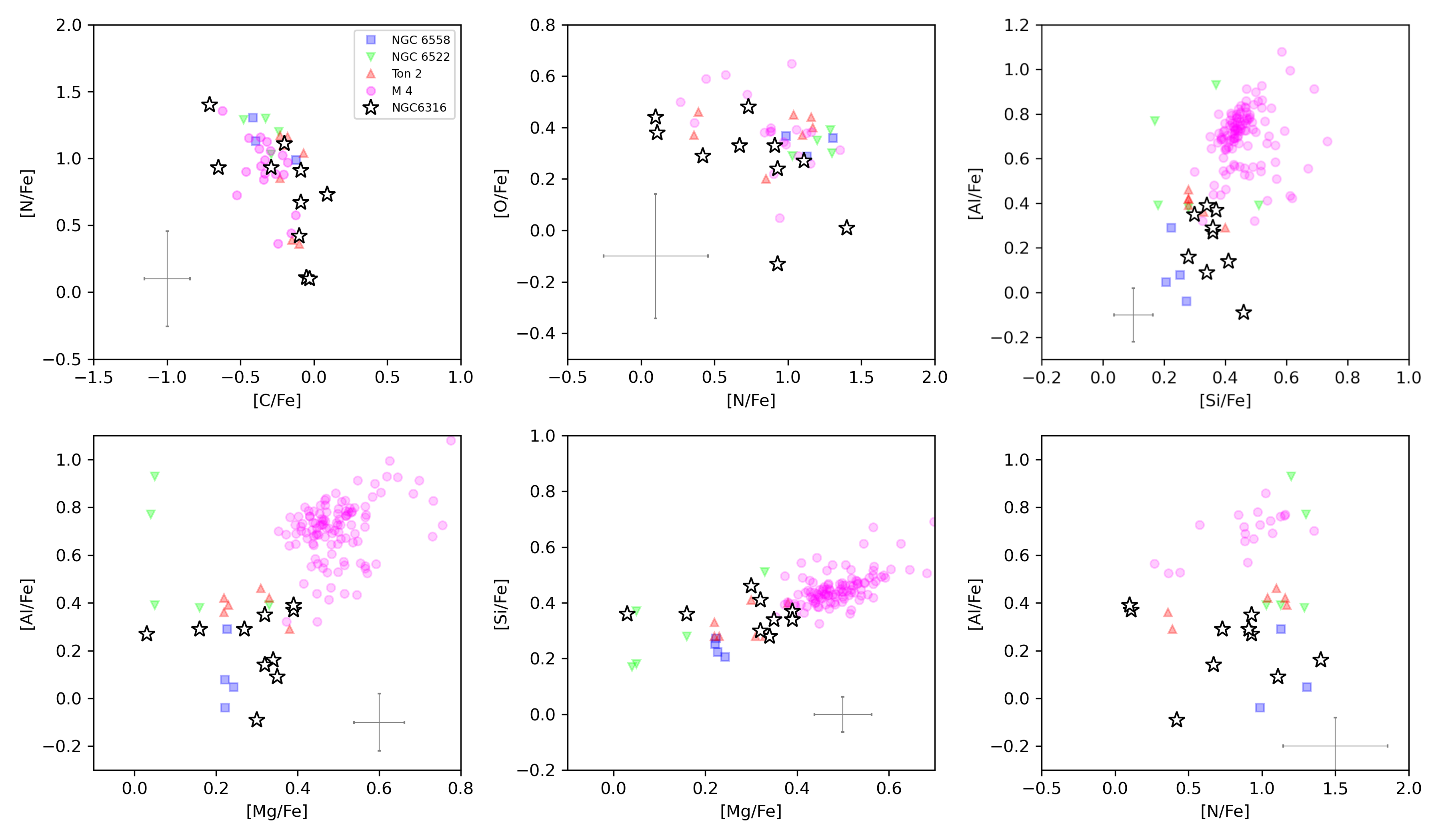}
    \caption{Abundance distributions of [N/Fe]-[C/Fe], [O/Fe]-[N/Fe], [Al/Fe]-[Si/Fe], [Al/Fe]-[Mg/Fe], [Si/Fe]-[Mg/Fe], and [Al/Fe]-[N/Fe] from our results for NGC 6316 are presented. The BGCs with similar metallicities, namely NGC 6558, NGC 6522, M4, and the disk GC Ton 2 were included for comparison. Associated mean errors for each abundance are shown in each plot.}
 	\label{Fabus}
\end{figure*}

Four different abundance determination methods are used, that are: line-profile fitting, an equivalent-width comparison, core line-intensity comparison and global goodness-of-fit estimate. For each method, we used a set of synthetic spectra of different abundances to compute the observed abundance. Only the lines with the best fit of the individual abundances are kept. We chose the $\chi^2$ goodness-of-fit diagnostic for the final abundance determination due to its reliability and consistency with our previous work. Nevertheless, we preserved data from the other diagnostic methods. To determine the C, N, and O elemental abundances, a mix of heavily CN-cycled and $\alpha$-rich models were used, as well as the molecular lines mentioned earlier. Oxygen abundances are estimated from the hydroxide molecular lines. Then, with these abundances in hand, we derived the carbon abundances from the carbon monoxide molecular lines, and last, the nitrogen abundances are obtained from the cyanogen molecular lines. We repeated this cycle until any dependence on OH, CO, and CN lines was negligible. After this process is finished, the rest of the mentioned elements are derived using the same four abundance determination methods.

Figure \ref{lines} shows the fit obtained for Oxygen, Aluminium and Nickel for the star 2M17163903$-$2807212 as an example of the quality of the fit described. In the Figure, the black dotted line is the observed spectrum of the target and the red continuum line is the created synthetic spectrum.

To determine and weight the errors we determined the abundances for each star again, varying only one of the parameters T$_{eff}$, log g, v$_t$ and [Fe/H] by $\pm$100 K, $\pm$0.3, $\pm$0.05 Km$\cdot$s$^{-1}$ and $\pm$0.05, respectively, for a total of 8 different abundances for each element. These
values were chosen as they represent the typical uncertainty in the atmospheric parameters for our sample \citep[see e.g.][]{Romero-Colmenares2021,GonzalezDiaz2023}. A standard deviation was calculated for each parameter separately ($\sigma_{T_{eff}}$, $\sigma_{logg}$, $\sigma{v_t}$ and $\sigma_{[Fe/H]}$), including the real abundance for each element with the 2 resulting abundances for the varied parameter, in the form: 

$${\rm\sigma_{T_{eff}} = \sigma\left([Fe/H],[Fe/H]^{T_{eff}}_{+100},[Fe/H]^{T_{eff}}_{-100}\right)}$$
$${\rm\sigma_{logg} = \sigma\left([ Fe/H],[Fe/H]^{logg}_{+0.3},[Fe/H]^{logg}_{-0.3}\right)}$$
$${\rm\sigma_{v_t} = \sigma\left([Fe/H],[Fe/H]^{v_t}_{+0.05},[Fe/H]^{v_t}_{-0.05}\right)}$$
$${\rm\sigma_{[Fe/H]} = \sigma \left([Fe/H],[Fe/H]^{[Fe/H]}_{+0.05},[Fe/H]^{[Fe/H]}_{-0.05}\right)}$$

The final error was determined by taking the square root of the quadratic sum of these four different errors:

$${\rm\sigma = \sqrt{{\sigma_{T_{eff}}}^2 + {\sigma_{logg}}^2 + {\sigma_{v_t}}^2 + {\sigma_{[Fe/H]}}^2}}$$

We finally took the errors from all the elements of all the stars and associated them with the corresponding abundances. 

\section{Results and discussion}
\label{Results}
\subsection{Results and comparison with other in-situ clusters}
We report abundances for C, N, O, Mg, Al, Si, P, K, Ca, Ti, V, Cr, Mn, Fe, Ni, and Ce. These are shown in Table \ref{abus} together with overall [$\alpha$/Fe] abundances for each star determined with the $\alpha$ elements Si and Ca, weighted equally. Each abundance has its associated weighted error, determined by the square root of the quadratic sum of each individual error and dividing the result by the number of stars. We adopted the solar reference abundances from \citet{Asplund2005} for all elements, as used in most CAPOS papers, due to their widespread adoption and consistency in the literature. Table \ref{abus} also contains the micro-turbulence velocities obtained by BACCHUS. This is the first spectroscopic report of abundances manually determined for this cluster and, since we used high-resolution spectroscopy, we believe the most robust study to date.
We obtain a mean metallicity for the cluster of [Fe/H] = $-$0.87 $\pm$ 0.02, with no evidence of any intrinsic Fe abundance variation. Our metallicity is in very good agreement with that obtained by D23, [Fe/H] = $-0.9$, and \citet{Conroy2018}, [Fe/H] = $-0.87$. However, it is some 0.5 dex lower than the canonical value of $-$0.45 from the H10 catalogue (that is a weighted average of the values derived by AZ88 scaled to C09, VFO07, and C09. The latter triple-weighted) and similarly offset from the \citet{Vasquez2018} values of $-0.35$, $-0.37$ and $-0.63$. As mentioned, these values were obtained by relatively uncertain photometric or low resolution spectroscopic means, and/or using a smaller number of samples. We also cannot discard uncertainties in the membership of some stars in their samples.
Thus, our metallicity determination should therefore be preferred over previous estimates, since it is based on a large sample of stars with a high probability of membership, observed at high resolution and SNR.

With our new metallicity, NGC 6316 is now no longer a member of the classic metal-rich BGCs. Indeed, based on both the \citet{Bica2024} and Geisler et al. (in prep) metallicity distribution functions of BGCs, which (assuming that it is indeed a BGC) are bimodal with peaks near -0.5 and -1.1, NGC 6316 falls in the lower metallicity domain, which is now known to be the dominant mode.

Given our good agreement with the D23 estimate, we note that this gives increased support to their age determination of 13.1 $\pm$ 0.5 Gyr, ensuring that NGC 6316 is indeed a rather old in situ GC for its metallicity. It now falls nicely along the age-metallicity sequence of in situ GCs, which are substantially more metal-rich than their accreted cousins of similar age, and indeed older than similar metallicity ex situ GCs \citep{Cohen2021,Belokurov2024}.
If instead the old H10 metallicity was used, NGC 6316 would be a substantial outlier in both of these diagrams.

We obtained an overall [$\alpha$/Fe] = 0.31 $\pm$ 0.02 determined using our silicon and calcium abundances. Oxygen and magnesium were excluded from the [$\alpha$/Fe] calculation, as these elements can vary substantially in globular clusters due to their involvement in MPs. To assess the reliability of our results, we determined the global metallicity [M/H] of the cluster using an alternative approach. Based on our derived [Fe/H] and [$\alpha$/Fe] abundances, we obtained an overall metallicity of [M/H]$=-0.65\pm0.08$ using the formula from \citet{Salaris1993}, in very good agreement with the value obtained through isochrone fitting ([M/H]$=-0.6\pm0.05$).

Figure \ref{Fabus} shows abundance relations for light elements involved in the MP phenomenon. Alongside NGC 6316 (white stars) there are four other clusters: NGC 6522 in emerald inverted triangles ([Fe/H] = $-1.04$, \citealt{Fernandez2019}), NGC 6558 in blue squares ([Fe/H] = $-1.15$, \citealt{GonzalezDiaz2023}), Ton 2 in red triangles ([Fe/H] = $-0.70$, \citealt{Fernandez2021}) and NGC 6121 (M4) in Fuchsia circles ([Fe/H] = $-1.02$, \citealt{Meszaros2020}. These clusters were selected because all of their abundances were determined using BACCHUS, and they present similar metallicities as our analysed stars. These criteria were selected in an attempt to find similar (or different) behaviours at similar/different location for in-situ GCs. According to Geisler et al. (in prep), all 5 of these GCs are in-situ, NGC 6316, NGC 6522 and NGC 6558 are indeed from the Bulge, Ton 2 is from the Disk, and it is uncertain whether NGC 6121 is from the Bulge or Disk.

As seen in Figure \ref{Fabus} top left panel, NGC 6316 presents a strong C-N anti-correlation, as do NGC 6522, NGC 6558, Ton 2 and NGC 6121, but our GC presents the strongest and most extended anti-correlation of all the compared B/D clusters.
 There is a significant spread in Al and Mg, with the latter spread being almost equal to M4 and higher than the other GCs compared in the plots. This spread in both elements is expected, as \citet{Carretta2009b} establishes that Al-rich and Mg-depleted stars are present in clusters that are massive (M $>$ $\sim10^5M_{\odot}$), quite metal-poor ([Fe/H] $< -1.5$), or both. Our cluster, although it is not considered "metal poor", has a mass of $3.47 \pm 0.44 \times 10^5 M_{\odot}$ according to the Baumgardt database, which is sufficient to be considered massive. However, the bottom left panel does not show a linear Mg-Al anti-correlation but rather a scatter in both elements.
\citet{Meszaros2020} defined 2P as [Al/Fe] $> +0.3$. However, this definition is not reliable for our case, since two of the three stars above this limit are the lowest in Nitrogen (hence deemed 1P) in our sample, using the definition of [N/Fe]$_{\rm 2P} >+0.7$ from \citet{Geisler2021}. It is worth noting that these two nitrogen-low stars are also the highest in magnesium in the sample (bottom left panel in Figure \ref{Fabus}). 
Taken as a whole, NGC 6316 displays most of the usual symptoms of the MP disease, but this is the first time it has been diagnosed in detail.

\begin{figure}
\centering
  	\includegraphics[width=1\columnwidth]{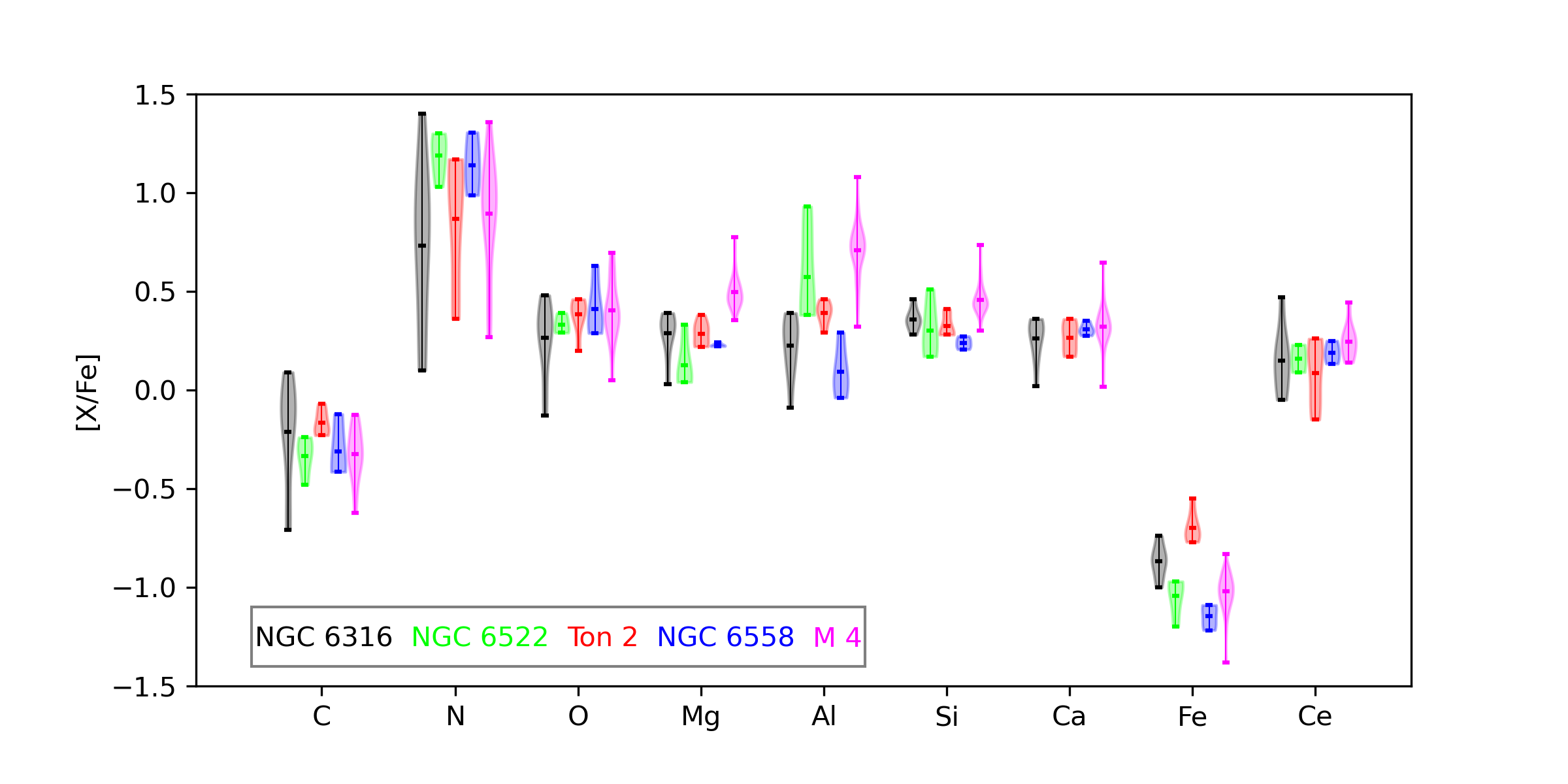}
    \caption{Violin diagram comparing the same GCs of Figure \ref{Fabus}. The vertical line indicate the spread in abundance for each element. And for each line, the most wide part indicates the major number of stars with the same abundance. For iron, interpret the y-axis as [Fe/H].}
 	\label{violin}
\end{figure}

Figure \ref{violin} shows a violin diagram comparing C, N, O, Mg, Al, Si, Ca, Fe, and Ce in NGC 6316 with those of the same clusters as in Figure \ref{Fabus}: NGC 6522, Ton 2, NGC 6558 and NGC 6121. This diagram illustrates the range for each element, represented by a vertical line with two small horizontal lines at the limits and a third one in-between showing the mean abundance of the element. The colour width of each element across the vertical line indicates the number of stars with each abundance, revealing how many stars contribute to the spread (the difference among the star with the highest abundance and that with the least abundance) seen in each element.

Our cluster shows a higher spread than the other in situ GCs in C, N and Ce, with only M4 surpassing it in O, Mg and Ca. Additionally, there is generally good agreement among these GC abundances and spreads, with the exception of N and Al.

Despite knowing that ASPCAP does not perform optimally for stars with extreme types of “non-standard” elemental abundance patterns, we took the opportunity to compare our abundances with the ASPCAP values provided in the APOGEE value-added catalogue from \citet{Schiavon2024}. This comparison is available in the appendix \ref{aspcap}.

\subsection{The in-situ nature of NGC 6316}

\begin{figure}
\centering
    \includegraphics[width=\linewidth]{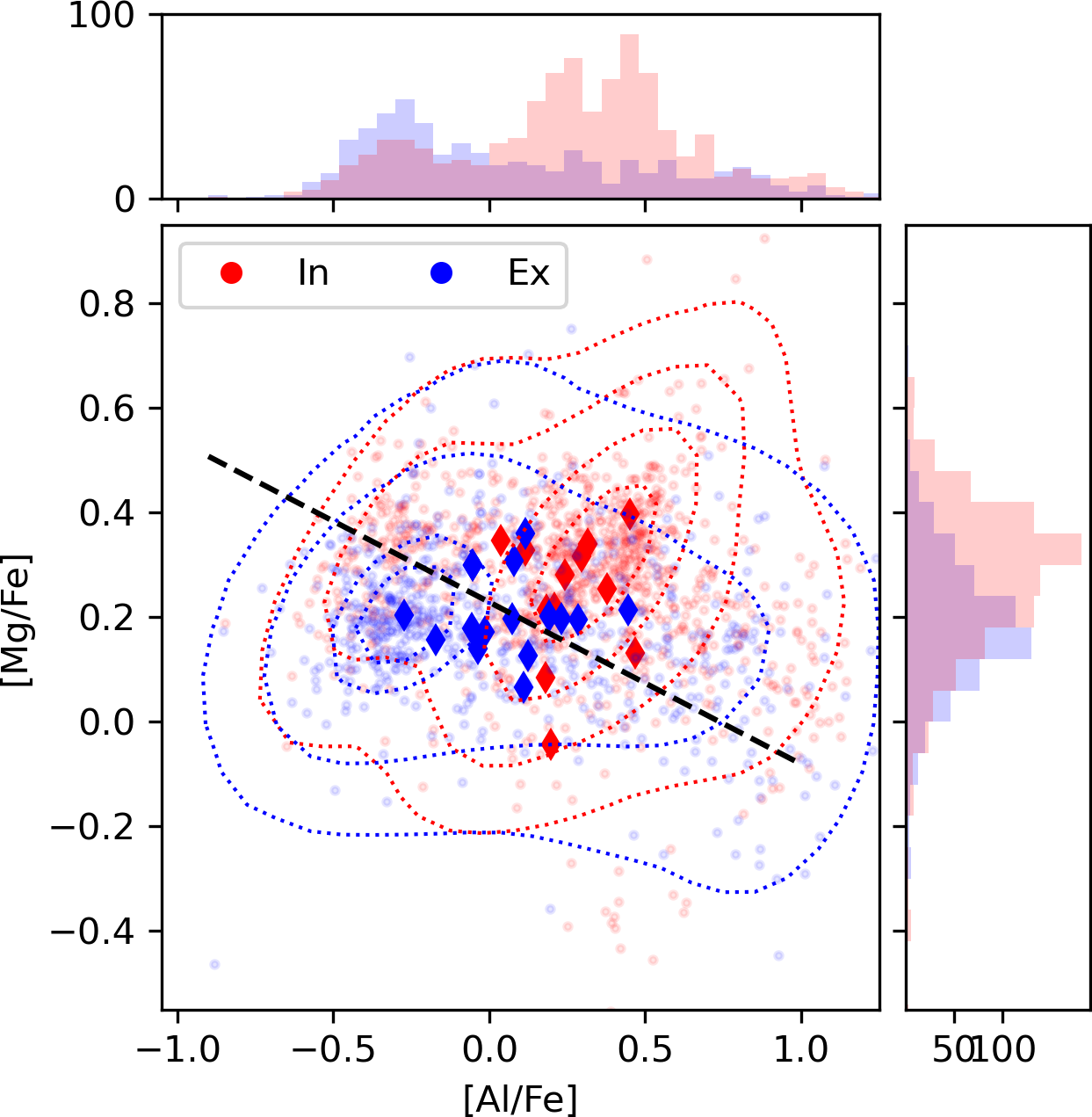}

    \caption{Mg vs Al-plane using BACCHUS abundances from \citet{Meszaros2020}. The abundances were systematically shifted as indicated in the appendix A of \citet{Belokurov2024}. Semi-transparent circles represent single stars, while diamonds represents mean abundances for each cluster. Colour dashed lines are the contours of the distribution of the single stars. The distribution of Mg and Al are shown in histograms at the right and top of the plot, respectively. Single star distributions show two clear dense areas where the in-and ex-situ stars are concentrated. This is more difficult to reproduce with mean values.}
 	\label{insitu}
\end{figure}
As mentioned, \citet{Massari2019} and Geisler et al. (in prep) classify NGC 6316 as an in-situ GC. \citet[hereafter BK24]{Belokurov2024} proposed an abundance-based classification scheme for in-situ and ex-situ GCs, utilizing [Mg/Fe] vs [Al/Fe] abundances derived from ASPCAP data and other sources scaled to ASPCAP (see their Fig. 11, right panel). We replicated this scheme using all GCs from \citet{Meszaros2020} with more than five stars, systematically shifting the data following the BK24 methodology (see their Appendix A) to align the [Mg/Fe] and [Al/Fe] abundances with the BK24 separation line distinguishing in-situ and ex-situ GCs (Figure \ref{insitu}).
In there, semi-transparent circles represent individual stars from \citet{Meszaros2020}, while diamonds indicate the mean abundance for each cluster. The colour-dashed lines outline the distribution contours of individual stars. The upper and right subplots display histograms of the overall dataset's [Mg/Fe] and [Al/Fe] distributions. The single-star distributions exhibit two distinct dense regions, corresponding to the concentrations of in-situ and ex-situ stars. However, this distinction is less clear in the mean values (diamond symbols), where the two groups overlap within a broader area.

We found that the method proposed by BK24 did not function as expected. The separation among in- and ex-situ GCs seems to be more complicated than simply using a line.

However, given the concentration of in- and ex- situ individual stars, it seems that a possible division can be made using a subgroup of stars, most probably 1P stars, that do not suffer from abundance variations.
This subject, will be further explored in greater detail in a forthcoming paper (Geisler et al. in prep) focusing on 1P stars from all CAPOS clusters.

\section{Conclusions}
\label{Conclusions}
In this paper, we have derived detailed abundances for a wide variety of elements for 10 RGB stars in the BGC NGC 6316 using the Brussels Automatic Code for Characterizing High accuracy Spectra (BACCHUS) based on CAPOS spectra. These stars were selected based on their strong affinity with the cluster's mean spatial, orbital, and chemical parameters (TR, PMs, RVs, and ASPCAP [Fe/H]). Based on our findings, we can state the following:
   \begin{enumerate}
      \item We obtained a mean metallicity of [Fe/H]= $-$0.87 $\pm$ 0.02 for the cluster, with no clear evidence of any intrinsic Fe abundance variation, in very good agreement with the results of D23. This value is $\sim$0.5 dex more metal poor than the values traditionally derived for this cluster \citep[and H10]{Vasquez2018}. We encourage the use of this new derived metallicity above the previous ones since our determination involves a large sample of stars observed at high resolution and SNR. Also, since our metallicity supports that derived by D23, the age derived by them (13.1 ± 0.5 Gyrs) is therefore supported by this study. 
      
      \item We provide abundances for C, N, O, Mg, Al, Si, P, K, Ca, Ti, V, Cr, Mn, Ni and Ce. This is the first non-automatic high resolution spectroscopic abundance determination for this cluster, hence the most reliable chemical study to date for this cluster, preferable to previous studies like \citet{Conroy2018}, who used stellar population models, respectively. Using the abundances of Si and Ca $\alpha$-elements we derived [$\alpha$/Fe] = 0.31 $\pm$ 0.02.

      \item The C and N levels of the stars show a clear anti-correlation, with a noticeable enhancement in Nitrogen, indicating the presence of MPs within the cluster.

      \item The isochrone fitting included in our method for determining the atmospheric parameters provided the following information for NGC 6316: A$_V$ = 1.91 $\pm$ 0.05, R$_V$ = 2.7 $\pm$ 0.1 (hence giving a higher E(B-V) = 0.71 than other studies that use R$_V$ = 3.1), (m-M)$_0$ = 15.32 $\pm$ 0.05 and [M/H] = $-$0.6 $\pm$ 0.05. This latter is in good agreement with our global metallicity derived from our [Fe/H] and [$\alpha$/Fe] abundances ([M/H] = $-$0.65 $\pm$ 0.08), using the formula from \citet{Salaris1993}.
      
      \item A comparison of the abundances of NGC 6316 with those of other in-situ GCs shows a very good agreement in almost all cases, with NGC 6316 having a higher spread in C, N and Ce.
      
      \item We found that the method proposed by BK24 to classify GCs based on their [Mg/Fe] vs. [Al/Fe] abundances did not function as expected. However, a classification using only 1P stars could work.

   \end{enumerate}

\begin{acknowledgements}
The authors acknowledge the referee's comments that substantially improved the content of this paper.
H.F. gratefully acknowledges the support provided by ALMA-ANID funds, No. 31230013.
D.G.D. gratefully acknowledges the support provided by ALMA-ANID funds, No. 31230017.
S.V. gratefully acknowledges the support provided by Fondecyt regular n. 1220264 and by the ANID BASAL projects FB210003.
D.G. gratefully acknowledges the support provided by Fondecyt regular n. 1220264.
D.G. also acknowledges financial support from the Direcci\'on de Investigaci\'on y Desarrollo de la Universidad de La Serena through the Programa de Incentivo a la Investigaci\'on de Acad\'emicos (PIA-DIDULS).
C.M. thanks the support provided by ANID GEMINI Postdoctorado No. 32230017.
The team acknowledges J.G. Fernandez-Trincado for his help and commentaries.
Funding for the Sloan Digital Sky Survey IV has been provided by the Alfred P. Sloan Foundation, the U.S. Department of Energy Office of Science, and the Participating 
Institutions. SDSS-IV acknowledges support and resources from the Center for High Performance Computing  at the University of Utah. The SDSS website is www.sdss4.org. 
SDSS-IV is managed by the Astrophysical Research Consortium for the Participating Institutions of the SDSS Collaboration including the Brazilian Participation Group, 
the Carnegie Institution for Science, Carnegie Mellon University, Center for Astrophysics | Harvard \& Smithsonian, the Chilean Participation Group, the French Participation Group, Instituto de Astrof\'isica de Canarias, The Johns Hopkins University, Kavli Institute for the Physics and Mathematics of the Universe (IPMU) / University of Tokyo, the Korean Participation Group, Lawrence Berkeley National Laboratory, Leibniz Institut f\"ur Astrophysik Potsdam (AIP),  Max-Planck-Institut 
f\"ur Astronomie (MPIA Heidelberg), Max-Planck-Institut f\"ur Astrophysik (MPA Garching), Max-Planck-Institut f\"ur Extraterrestrische Physik (MPE), National Astronomical Observatories of China, New Mexico State University, New York University, University of Notre Dame, Observat\'ario Nacional / MCTI, The Ohio State 
University, Pennsylvania State University, Shanghai Astronomical Observatory, United 
Kingdom Participation Group, Universidad Nacional Aut\'onoma de M\'exico, University of Arizona, University of Colorado Boulder, University of Oxford, University of 
Portsmouth, University of Utah, University of Virginia, University of Washington, University of Wisconsin, Vanderbilt University, and Yale University.
This work has made use of data from the European Space Agency (ESA) mission Gaia (\url{https://www.cosmos.esa.int/gaia}), processed by the Gaia Data Processing and Analysis Consortium (DPAC, \url{https://www.cosmos.esa.int/web/gaia/dpac/consortium}). Funding for the DPAC has been provided by national institutions, in particular the institutions participating in the Gaia Multilateral Agreement.
\end{acknowledgements}

\begin{appendix}
\section{Comparisons with ASPCAP}
\label{aspcap}
Given the substantial number of new abundances obtained, it is a good opportunity to compare them with the ASPCAP values provided in the APOGEE value-added catalogue from \citet[hereafter S24]{Schiavon2024}. A similar analysis, expanded for all CAPOS cluster, is being carried out in Geisler et al. (in prep).
S24 presented abundances for a group of stars in NGC 6316, which includes all 10 of our target stars, enabling a reliable comparison. For brevity, we will refer to our results as BACCHUS and those from S24 as ASPCAP.

The mean iron abundance in BACCHUS ($-$0.87 $\pm$ 0.02) is approximately 0.1 dex lower than ASPCAP ($-$0.78 $\pm$ 0.005), yet still aligns with the metal-poor side of the measurements when comparing with the old references. ASPCAP also presents a lower metallicity dispersion than BACCHUS.
One of the most notable differences between the ASPCAP and BACCHUS abundances lies in Carbon. Although both methods show the same anti-correlation in C and N and similar trend, the two most Carbon-depleted stars, identified as 2P stars following the [N/Fe]$_{2P}>+0.7$ criteria from \citet{Geisler2021}, are 0.3-0.4 dex more depleted in BACCHUS than in ASPCAP. For Nitrogen, while ASPCAP and BACCHUS exhibit similar dispersion, ASPCAP reports a mean nitrogen abundance that is $\sim$0.1 dex lower than BACCHUS. Regarding the two mentioned 2P stars, their mean nitrogen values are similar in both methods, but the BACCHUS abundances show significantly higher dispersion.

\begin{figure}[ht]
\centering
  	\includegraphics[width=1\linewidth]{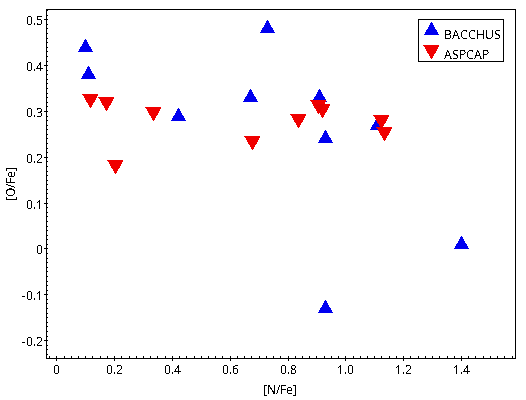}
    \caption{Comparison between Oxygen and Nitrogen values of ASPCAP and BACCHUS for the same 10 stars. Blue triangles indicate BACCHUS abundances while inverted red triangles indicate ASPCAP values. The N-O anti-correlation is clearer for BACCHUS abundances.}
 	\label{vs}
\end{figure}

The $\alpha$-elements Si and Ca show a difference of $\sim$0.2 in their overall [$\alpha$/Fe] abundance, with BACCHUS reporting a mean of [$\alpha$/Fe] = 0.31 $\pm$ 0.02, while ASPCAP reports a mean [$\alpha$/Fe] = 0.13. Oxygen shows similar mean values between ASPCAP ([O/Fe] = 0.28) and BACCHUS ([O/Fe] = 0.26), but BACCHUS exhibits significantly higher dispersion (0.04 vs 0.18). Since MPs are present in this cluster, an oxygen dispersion is expected, lending more credibility to our individual oxygen abundances. Indeed, a N-O anti-correlation (or C-O correlation) is not observed using ASPCAP values (Figure \ref{vs}).

Magnesium shows a strong agreement in its mean abundance between the two methods, but BACCHUS again exhibits higher dispersion. Notably, two stars in BACCHUS are more magnesium-depleted than any ASPCAP star. Among these, one also displays the high carbon depletion and enhanced nitrogen mentioned earlier, while the other presents a normal level of carbon but enhanced nitrogen abundance (and thus both considered as 2P). Aluminium and Potassium show excellent agreement in both their mean abundance and dispersion across methods. Both ASPCAP and BACCHUS report the same dispersion for Silicon, but BACCHUS abundances are $\sim$0.1 dex higher.
Calcium in BACCHUS has a slightly higher abundance, with dispersion also being twice as large as in ASPCAP. Titanium shows similar dispersion in both methods but ASPCAP reports abundances that are $\sim$0.05 dex higher. However, the typical errors of our Ti measurements allow to determine an agreement between ASPCAP and BACCHUS for this element.

The iron-peak elements chromium, manganese and nickel are in excellent agreement between both methods, both in terms of abundance and dispersion. Vanadium in BACCHUS is 0.2 dex higher than in ASPCAP, though its dispersion is twice as large in the latter. Cerium shows half the dispersion and abundance in ASPCAP compared to BACCHUS, but both methods reveal the correlation with nitrogen. 

It is worth noting that Vanadium is the only element that presents higher dispersion in ASPCAP abundances than in BACCHUS. 7 elements (C, N, O, Mg, Ca, Fe and Ce) present lower dispersion with the first method, while the other 7 elements (Al, Si, K, Ti, Cr, Mn and Ni) present the same dispersion with both methods. No P abundances are given by ASPCAP.

It should be noted that using our [Fe/H] and [$\alpha$/Fe] abundances to calculate the global metallicity [M/H] via the formula from \citet{Salaris1993}, we obtain a [M/H]$\sim$0.3 dex higher than our [Fe/H]. This is expected since GCs are usually $\alpha$-enhanced. Conversely, the global metallicity reported by ASPCAP is almost identical to their iron metallicity, differing by only $\sim$0.01 dex, which does not align with their calculation using the aforementioned formula. 
\end{appendix}

\begin{thebibliography}{99}
\bibitem[\protect\citeauthoryear{Abdurro'uf et al.}{2022}]{Abdurro'uf2022} Abdurro'uf, Accetta K., Aerts C., Silva Aguirre V., Ahumada R., Ajgaonkar N., Filiz Ak N., et al., 2022, ApJS, 259, 35. doi:10.3847/1538-4365/ac4414
\bibitem[\protect\citeauthoryear{Alvarez \& Plez}{1998}]{Alvarez1998} Alvarez R., Plez B., 1998, A\&A, 330, 1109. doi:10.48550/arXiv.astro-ph/9710157
\bibitem[\protect\citeauthoryear{Armandroff \& Zinn}{1988}]{Armandroff1988} Armandroff T.~E., Zinn R., 1988, AJ, 96, 92. doi:10.1086/114792
\bibitem[\protect\citeauthoryear{Asplund, Grevesse, \& Sauval}{2005}]{Asplund2005} Asplund M., Grevesse N., Sauval A.~J., 2005, ASPC, 336, 25
\bibitem[\protect\citeauthoryear{Barrera et al.}{2025}]{Barrera2025} Barrera N., Villanova S., Geisler D., Fern{\'a}ndez-Trincado J.~G., Mu{\~n}oz C., 2025, A\&A, 699, A128. doi:10.1051/0004-6361/202451968\bibitem[\protect\citeauthoryear{Belokurov \& Kravtsov}{2024}]{Belokurov2024} Belokurov V., Kravtsov A., 2024, MNRAS, 528, 3198. doi:10.1093/mnras/stad3920
\bibitem[\protect\citeauthoryear{Bica et al.}{2024}]{Bica2024} Bica E., Ortolani S., Barbuy B., Oliveira R.~A.~P., 2024, A\&A, 687, A201. doi:10.1051/0004-6361/202346377
\bibitem[\protect\citeauthoryear{Bowen \& Vaughan}{1973}]{Bowen1973} Bowen I.~S., Vaughan A.~H., 1973, ApOpt, 12, 1430. doi:10.1364/AO.12.001430
\bibitem[\protect\citeauthoryear{Bressan et al.}{2012}]{Bressan2012} Bressan A., Marigo P., Girardi L., Salasnich B., Dal Cero C., Rubele S., Nanni A., 2012, MNRAS, 427, 127. doi:10.1111/j.1365-2966.2012.21948.x
\bibitem[\protect\citeauthoryear{Callingham et al.}{2022}]{Callingham2022} Callingham T.~M., Cautun M., Deason A.~J., Frenk C.~S., Grand R.~J.~J., Marinacci F., 2022, MNRAS, 513, 4107. doi:10.1093/mnras/stac1145
\bibitem[\protect\citeauthoryear{Caloi \& D'Antona}{2008}]{Caloi2008} Caloi V., D'Antona F., 2008, ApJ, 673, 847. doi:10.1086/523346
\bibitem[\protect\citeauthoryear{Cardelli, Clayton, \& Mathis}{1989}]{Cardelli1989} Cardelli J.~A., Clayton G.~C., Mathis J.~S., 1989, ApJ, 345, 245. doi:10.1086/167900
\bibitem[\protect\citeauthoryear{Carretta et al.}{2009a}]{Carretta2009a} Carretta E., Bragaglia A., Gratton R.~G., Lucatello S., Catanzaro G., Leone F., Bellazzini M., et al., 2009, A\&A, 505, 117. doi:10.1051/0004-6361/200912096
\bibitem[\protect\citeauthoryear{Carretta et al.}{2009b}]{Carretta2009b} Carretta E., Bragaglia A., Gratton R., Lucatello S., 2009, A\&A, 505, 139. doi:10.1051/0004-6361/200912097
\bibitem[\protect\citeauthoryear{Carretta et al.}{2009c}]{Carretta2009c} Carretta E., Bragaglia A., Gratton R., D'Orazi V., Lucatello S., 2009, A\&A, 508, 695. doi:10.1051/0004-6361/200913003
\bibitem[\protect\citeauthoryear{Chen \& Gnedin}{2024}]{Chen2024} Chen Y., Gnedin O.~Y., 2024, MNRAS, 527, 3692. doi:10.1093/mnras/stad3345
\bibitem[\protect\citeauthoryear{Cohen \& Sarajedini}{2012}]{Cohen2012} Cohen R.~E., Sarajedini A., 2012, MNRAS, 419, 342. doi:10.1111/j.1365-2966.2011.19697.x
\bibitem[\protect\citeauthoryear{Cohen et al.}{2021}]{Cohen2021} Cohen R.~E., Bellini A., Casagrande L., Brown T.~M., Correnti M., Kalirai J.~S., 2021, AJ, 162, 228. doi:10.3847/1538-3881/ac281f
\bibitem[\protect\citeauthoryear{Conroy et al.}{2018}]{Conroy2018} Conroy C., Villaume A., van Dokkum P.~G., Lind K., 2018, ApJ, 854, 139. doi:10.3847/1538-4357/aaab49
\bibitem[\protect\citeauthoryear{Cunha et al.}{2017}]{Cunha2017} Cunha K., Smith V.~V., Hasselquist S., Souto D., Shetrone M.~D., Allende Prieto C., Bizyaev D., et al., 2017, ApJ, 844, 145. doi:10.3847/1538-4357/aa7beb
\bibitem[\protect\citeauthoryear{Deras et al.}{2023}]{Deras2023} Deras D., Cadelano M., Ferraro F.~R., Lanzoni B., Pallanca C., 2023, ApJ, 942, 104. doi:10.3847/1538-4357/aca9ce
\bibitem[\protect\citeauthoryear{De Silva et al.}{2015}]{Desilva2015} De Silva G.~M., Freeman K.~C., Bland-Hawthorn J., Martell S., de Boer E.~W., Asplund M., Keller S., et al., 2015, MNRAS, 449, 2604. doi:10.1093/mnras/stv327
\bibitem[\protect\citeauthoryear{Dias et al.}{2016}]{Dias2016} Dias B., Barbuy B., Saviane I., Held E.~V., Da Costa G.~S., Ortolani S., Gullieuszik M., et al., 2016, A\&A, 590, A9. doi:10.1051/0004-6361/201526765
\bibitem[\protect\citeauthoryear{Dotter et al.}{2010}]{Dotter2010} Dotter A., Sarajedini A., Anderson J., Aparicio A., Bedin L.~R., Chaboyer B., Majewski S., et al., 2010, ApJ, 708, 698. doi:10.1088/0004-637X/708/1/698
\bibitem[\protect\citeauthoryear{Fern{\'a}ndez-Trincado et al.}{2019}]{Fernandez2019} Fern{\'a}ndez-Trincado J.~G., Zamora O., Souto D., Cohen R.~E., Dell'Agli F., Garc{\'\i}a-Hern{\'a}ndez D.~A., Masseron T., et al., 2019, A\&A, 627, A178. doi:10.1051/0004-6361/201834391
\bibitem[\protect\citeauthoryear{Fern{\'a}ndez-Trincado et al.}{2020}]{Fernandez2020} Fern{\'a}ndez-Trincado J.~G., Beers T.~C., Minniti D., Tang B., Villanova S., Geisler D., P{\'e}rez-Villegas A., et al., 2020, A\&A, 643, L4. doi:10.1051/0004-6361/202039207
\bibitem[\protect\citeauthoryear{Fern{\'a}ndez-Trincado et al.}{2021}]{Fernandez2021} Fern{\'a}ndez-Trincado J.~G., Beers T.~C., Barbuy B., M{\'e}sz{\'a}ros S., Minniti D., Smith V.~V., Cunha K., et al., 2021, ApJL, 918, L9. doi:10.3847/2041-8213/ac1c7e
\bibitem[\protect\citeauthoryear{Fern{\'a}ndez-Trincado et al.}{2022}]{Fernandez2022} Fern{\'a}ndez-Trincado J.~G., Villanova S., Geisler D., Barbuy B., Minniti D., Beers T.~C., M{\'e}sz{\'a}ros S., et al., 2022, A\&A, 658, A116. doi:10.1051/0004-6361/202141742
\bibitem[\protect\citeauthoryear{Frelijj et al.}{2021}]{Frelijj2021} Frelijj H., Villanova S., Mu{\~n}oz C., Fern{\'a}ndez-Trincado J.~G., 2021, MNRAS, 503, 867. doi:10.1093/mnras/stab443
\bibitem[\protect\citeauthoryear{Gaia Collaboration et al.}{2016}]{Gaia2016} Gaia Collaboration, Prusti T., de Bruijne J.~H.~J., Brown A.~G.~A., Vallenari A., Babusiaux C., Bailer-Jones C.~A.~L., et al., 2016, A\&A, 595, A1. doi:10.1051/0004-6361/201629272
\bibitem[\protect\citeauthoryear{Gaia Collaboration et al.}{2021}]{Gaia2021} Gaia Collaboration, Brown A.~G.~A., Vallenari A., Prusti T., de Bruijne J.~H.~J., Babusiaux C., Biermann M., et al., 2021, A\&A, 649, A1. doi:10.1051/0004-6361/202039657
\bibitem[\protect\citeauthoryear{Garc{\'\i}a P{\'e}rez et al.}{2016}]{Garcia2016} Garc{\'\i}a P{\'e}rez A.~E., Allende Prieto C., Holtzman J.~A., Shetrone M., M{\'e}sz{\'a}ros S., Bizyaev D., Carrera R., et al., 2016, AJ, 151, 144. doi:10.3847/0004-6256/151/6/144
\bibitem[\protect\citeauthoryear{Geisler et al.}{2021}]{Geisler2021} Geisler D., Villanova S., O'Connell J.~E., Cohen R.~E., Moni Bidin C., Fern{\'a}ndez-Trincado J.~G., Mu{\~n}oz C., et al., 2021, A\&A, 652, A157. doi:10.1051/0004-6361/202140436
\bibitem[\protect\citeauthoryear{Goldsbury, Heyl, \& Richer}{2013}]{Goldsbury2013} Goldsbury R., Heyl J., Richer H., 2013, ApJ, 778, 57. doi:10.1088/0004-637X/778/1/57
\bibitem[\protect\citeauthoryear{Gonzalez et al.}{2012}]{Gonzalez2012} Gonzalez O.~A., Rejkuba M., Zoccali M., Valenti E., Minniti D., Schultheis M., Tobar R., et al., 2012, A\&A, 543, A13. doi:10.1051/0004-6361/201219222
\bibitem[\protect\citeauthoryear{Gonz{\'a}lez-D{\'\i}az et al.}{2023}]{GonzalezDiaz2023} Gonz{\'a}lez-D{\'\i}az D., Fern{\'a}ndez-Trincado J.~G., Villanova S., Geisler D., Barbuy B., Minniti D., Beers T.~C., et al., 2023, MNRAS, 526, 6274. doi:10.1093/mnras/stad3178
\bibitem[\protect\citeauthoryear{Gustafsson et al.}{2008}]{Gustafsson2008} Gustafsson B., Edvardsson B., Eriksson K., J{\o}rgensen U.~G., Nordlund {\r{A}}., Plez B., 2008, A\&A, 486, 951. doi:10.1051/0004-6361:200809724
\bibitem[\protect\citeauthoryear{Harris}{1996}]{Harris1996} Harris W.~E., 1996, AJ, 112, 1487. doi:10.1086/118116
\bibitem[\protect\citeauthoryear{Hasselquist et al.}{2016}]{Hasselquist2016} Hasselquist S., Shetrone M., Cunha K., Smith V.~V., Holtzman J., Lawler J.~E., Allende Prieto C., et al., 2016, ApJ, 833, 81. doi:10.3847/1538-4357/833/1/81
\bibitem[\protect\citeauthoryear{Hawkins et al.}{2016}]{Hawkins2016} Hawkins K., Masseron T., Jofr{\'e} P., Gilmore G., Elsworth Y., Hekker S., 2016, A\&A, 594, A43. doi:10.1051/0004-6361/201628812
\bibitem[\protect\citeauthoryear{Henao et al.}{2025}]{Henao2025} Henao L., Villanova S., Geisler D., Fern{\'a}ndez-Trincado J.~G., 2025, A\&A, 696, A154. doi:10.1051/0004-6361/202451793
\bibitem[\protect\citeauthoryear{Holtzman et al.}{2015}]{Holtzman2015} Holtzman J.~A., Shetrone M., Johnson J.~A., Allende Prieto C., Anders F., Andrews B., Beers T.~C., et al., 2015, AJ, 150, 148. doi:10.1088/0004-6256/150/5/148
\bibitem[\protect\citeauthoryear{J{\"o}nsson et al.}{2018}]{Jonsson2018} J{\"o}nsson H., Allende Prieto C., Holtzman J.~A., Feuillet D.~K., Hawkins K., Cunha K., M{\'e}sz{\'a}ros S., et al., 2018, AJ, 156, 126. doi:10.3847/1538-3881/aad4f5
\bibitem[\protect\citeauthoryear{Majewski et al.}{2017}]{Majewski2017} Majewski S.~R., Schiavon R.~P., Frinchaboy P.~M., Allende Prieto C., Barkhouser R., Bizyaev D., Blank B., et al., 2017, AJ, 154, 94. doi:10.3847/1538-3881/aa784d
\bibitem[\protect\citeauthoryear{Massari, Koppelman, \& Helmi}{2019}]{Massari2019} Massari D., Koppelman H.~H., Helmi A., 2019, A\&A, 630, L4. doi:10.1051/0004-6361/201936135
\bibitem[\protect\citeauthoryear{Masseron, Merle, \& Hawkins}{2016}]{Masseron2016} Masseron T., Merle T., Hawkins K., 2016, ascl.soft. ascl:1605.004
\bibitem[\protect\citeauthoryear{M{\'e}sz{\'a}ros et al.}{2020}]{Meszaros2020} M{\'e}sz{\'a}ros S., Masseron T., Garc{\'\i}a-Hern{\'a}ndez D.~A., Allende Prieto C., Beers T.~C., Bizyaev D., Chojnowski D., et al., 2020, MNRAS, 492, 1641. doi:10.1093/mnras/stz3496
\bibitem[\protect\citeauthoryear{Milone et al.}{2017}]{Milone2017} Milone A.~P., Piotto G., Renzini A., Marino A.~F., Bedin L.~R., Vesperini E., D'Antona F., et al., 2017, MNRAS, 464, 3636. doi:10.1093/mnras/stw2531
\bibitem[\protect\citeauthoryear{Minniti et al.}{2010}]{Minniti2010} Minniti D., Lucas P.~W., Emerson J.~P., Saito R.~K., Hempel M., Pietrukowicz P., Ahumada A.~V., et al., 2010, NewA, 15, 433. doi:10.1016/j.newast.2009.12.002
\bibitem[\protect\citeauthoryear{Momany et al.}{2012}]{Momany2012} Momany Y., Saviane I., Smette A., Bayo A., Girardi L., Marconi G., Milone A.~P., et al., 2012, A\&A, 537, A2. doi:10.1051/0004-6361/201117223
\bibitem[\protect\citeauthoryear{Mu{\~n}oz et al.}{2017}]{Munoz2017} Mu{\~n}oz C., Villanova S., Geisler D., Saviane I., Dias B., Cohen R.~E., Mauro F., 2017, A\&A, 605, A12. doi:10.1051/0004-6361/201730468
\bibitem[\protect\citeauthoryear{Mu{\~n}oz et al.}{2020}]{Munoz2020} Mu{\~n}oz C., Villanova S., Geisler D., Cort{\'e}s C.~C., Moni Bidin C., Cohen R.~E., Saviane I., et al., 2020, MNRAS, 492, 3742. doi:10.1093/mnras/stz3586
\bibitem[\protect\citeauthoryear{Nataf et al.}{2013}]{Nataf2013} Nataf D.~M., Gould A., Fouqu{\'e} P., Gonzalez O.~A., Johnson J.~A., Skowron J., Udalski A., et al., 2013, ApJ, 769, 88. doi:10.1088/0004-637X/769/2/88
\bibitem[\protect\citeauthoryear{Nidever et al.}{2015}]{Nidever2015} Nidever D.~L., Holtzman J.~A., Allende Prieto C., Beland S., Bender C., Bizyaev D., Burton A., et al., 2015, AJ, 150, 173. doi:10.1088/0004-6256/150/6/173
\bibitem[\protect\citeauthoryear{Origlia et al.}{2005}]{Origlia2005} Origlia L., Valenti E., Rich R.~M., Ferraro F.~R., 2005, MNRAS, 363, 897. doi:10.1111/j.1365-2966.2005.09490.x
\bibitem[\protect\citeauthoryear{Paust et al.}{2010}]{Paust2010} Paust N.~E.~Q., Reid I.~N., Piotto G., Aparicio A., Anderson J., Sarajedini A., Bedin L.~R., et al., 2010, AJ, 139, 476. doi:10.1088/0004-6256/139/2/476
\bibitem[\protect\citeauthoryear{P{\'e}rez-Villegas et al.}{2020}]{Perez2020} P{\'e}rez-Villegas A., Barbuy B., Kerber L.~O., Ortolani S., Souza S.~O., Bica E., 2020, MNRAS, 491, 3251. doi:10.1093/mnras/stz3162
\bibitem[\protect\citeauthoryear{Piotto et al.}{2015}]{Piotto2015} Piotto G., Milone A.~P., Bedin L.~R., Anderson J., King I.~R., Marino A.~F., Nardiello D., et al., 2015, AJ, 149, 91. doi:10.1088/0004-6256/149/3/91
\bibitem[\protect\citeauthoryear{Plez}{2012}]{Plez2012} Plez B., 2012, ascl.soft. ascl:1205.004
\bibitem[\protect\citeauthoryear{Reed, Hesser, \& Shawl}{1988}]{Reed1988} Reed B.~C., Hesser J.~E., Shawl S.~J., 1988, PASP, 100, 545. doi:10.1086/132202
\bibitem[\protect\citeauthoryear{Romero-Colmenares et al.}{2021}]{Romero-Colmenares2021} Romero-Colmenares M., Fern{\'a}ndez-Trincado J.~G., Geisler D., Souza S.~O., Villanova S., Longa-Pe{\~n}a P., Minniti D., et al., 2021, A\&A, 652, A158. doi:10.1051/0004-6361/202141294
\bibitem[\protect\citeauthoryear{Saito et al.}{2024}]{Saito2024} Saito R.~K., Hempel M., Alonso-Garc{\'\i}a J., Lucas P.~W., Minniti D., Alonso S., Baravalle L., et al., 2024, A\&A, 689, A148. doi:10.1051/0004-6361/202450584\bibitem[\protect\citeauthoryear{Salaris, Chieffi, \& Straniero}{1993}]{Salaris1993} Salaris M., Chieffi A., Straniero O., 1993, ApJ, 414, 580. doi:10.1086/173105
\bibitem[\protect\citeauthoryear{Santana et al.}{2021}]{Santana2021} Santana F.~A., Beaton R.~L., Covey K.~R., O'Connell J.~E., Longa-Pe{\~n}a P., Cohen R., Fern{\'a}ndez-Trincado J.~G., et al., 2021, AJ, 162, 303. doi:10.3847/1538-3881/ac2cbc
\bibitem[\protect\citeauthoryear{Sarajedini et al.}{2007}]{Sarajedini2007} Sarajedini A., Bedin L.~R., Chaboyer B., Dotter A., Siegel M., Anderson J., Aparicio A., et al., 2007, AJ, 133, 1658. doi:10.1086/511979
\bibitem[\protect\citeauthoryear{Saviane et al.}{2012}]{Saviane2012} Saviane I., Da Costa G.~S., Held E.~V., Sommariva V., Gullieuszik M., Barbuy B., Ortolani S., 2012, A\&A, 540, A27. doi:10.1051/0004-6361/201118138
\bibitem[\protect\citeauthoryear{Schiavon et al.}{2024}]{Schiavon2024} Schiavon R.~P., Phillips S.~G., Myers N., Horta D., Minniti D., Allende Prieto C., Anguiano B., et al., 2024, MNRAS, 528, 1393. doi:10.1093/mnras/stad3020
\bibitem[\protect\citeauthoryear{Shetrone et al.}{2015}]{Shetrone2015} Shetrone M., Bizyaev D., Lawler J.~E., Allende Prieto C., Johnson J.~A., Smith V.~V., Cunha K., et al., 2015, ApJS, 221, 24. doi:10.1088/0067-0049/221/2/24
\bibitem[\protect\citeauthoryear{Valenti, Ferraro, \& Origlia}{2007}]{Valenti2007} Valenti E., Ferraro F.~R., Origlia L., 2007, AJ, 133, 1287. doi:10.1086/511271
\bibitem[\protect\citeauthoryear{Valenti, Origlia, \& Rich}{2011}]{Valenti2011} Valenti E., Origlia L., Rich R.~M., 2011, MNRAS, 414, 2690. doi:10.1111/j.1365-2966.2011.18580.x
\bibitem[\protect\citeauthoryear{VandenBerg et al.}{2013}]{Vandenberg2013} VandenBerg D.~A., Brogaard K., Leaman R., Casagrande L., 2013, ApJ, 775, 134. doi:10.1088/0004-637X/775/2/134
\bibitem[\protect\citeauthoryear{Vasiliev \& Baumgardt}{2021}]{Vasiliev2021} Vasiliev E., Baumgardt H., 2021, yCat, 750. J/MNRAS/505/5978
\bibitem[\protect\citeauthoryear{V{\'a}squez et al.}{2015}]{Vasquez2015} V{\'a}squez S., Zoccali M., Hill V., Gonzalez O.~A., Saviane I., Rejkuba M., Battaglia G., 2015, A\&A, 580, A121. doi:10.1051/0004-6361/201526534
\bibitem[\protect\citeauthoryear{V{\'a}squez et al.}{2018}]{Vasquez2018} V{\'a}squez S., Saviane I., Held E.~V., Da Costa G.~S., Dias B., Gullieuszik M., Barbuy B., et al., 2018, A\&A, 619, A13. doi:10.1051/0004-6361/201833525
\bibitem[Villanova et al.(2013)]{Villanova2013} Villanova, S., Geisler, D., Carraro, G., Moni Bidin, C., \& Mu{\~n}oz, C.\ 2013, \apj, 778, 186 
\bibitem[\protect\citeauthoryear{Villanova et al.}{2017}]{Villanova2017} Villanova S., Moni Bidin C., Mauro F., Munoz C., Monaco L., 2017, MNRAS, 464, 2730. doi:10.1093/mnras/stw2509
\bibitem[\protect\citeauthoryear{Wilson et al.}{2019}]{Wilson2019} Wilson J.~C., Hearty F.~R., Skrutskie M.~F., Majewski S.~R., Holtzman J.~A., Eisenstein D., Gunn J., et al., 2019, PASP, 131, 055001. doi:10.1088/1538-3873/ab0075
\bibitem[\protect\citeauthoryear{Zinn \& West}{1984}]{Zinn1984} Zinn R., West M.~J., 1984, ApJS, 55, 45. doi:10.1086/190947
\bibitem[\protect\citeauthoryear{Zinn}{1985}]{Zinn1985} Zinn R., 1985, ApJ, 293, 424. doi:10.1086/163249

\end{thebibliography}
\end{document}